\documentclass[%
 reprint,
 amsmath,amssymb,
 aps,
]{revtex4-2}

\usepackage{graphicx}
\usepackage{dcolumn}
\usepackage{bm}
\usepackage{multirow}
\usepackage{xcolor}
\usepackage{hyperref}

\begin{document}

\title{Generalized Radial Uncertainty Product for d-Dimensional Hydrogen Atom}

\author{Avoy Jana}
\affiliation{%
 Master in Science, Physics\\
 Indian Institute of Technology, Delhi
}%

\date{\today}

\begin{abstract}
\textbf{Abstract} This paper presents a comprehensive analysis of the generalized radial uncertainty product for the d-dimensional non-relativistic Hydrogen atom in position space. Utilizing the framework of quantum mechanics in d-dimensional spherical coordinates, the study extends the standard radial uncertainty relation to higher dimensions. Taking the solution of the radial Schrödinger equation, the normalized radial wave functions, expectation values, and uncertainties in both position and momentum space are rigorously evaluated. The analytical derivations reveal the dependence of the uncertainty product on the principal and angular quantum numbers, as well as the dimensional parameter d. The results provide deeper insight into the role of dimensionality in quantum uncertainty relations and their implications for higher-dimensional quantum systems.\\

\end{abstract}

\maketitle

\section{Introduction}
We have searched (Ref.~\cite{jana2025}) the radial uncertainty product $\Delta \hat{r}\Delta \hat p_r$ for some non-relativistic spherically symmetric potentials including Hydrogen atom. In three dimensional system, it's not a problem when we are evaluating the squared radial momentum expectation value using the total energy formula. Meanwhile the connection (Ref.~\cite{Paz2001}) between the Hamiltonian and the radial momentum operator,  
\begin{equation*}
    \hat{H} = \frac{\hat{p}_r^2}{2\mu} + \frac{\hat{L}^2}{2\mu r^2}
\end{equation*}
is valid only in one or three spatial dimensions. Then for an arbitrary dimensional Hydrogen atom, we have to sure about the true connection between them. In many literature, the uncertainty relation is shown as the product of the expectation values of the squared position operator, $\langle \hat r^2 \rangle$ (Ref.~\cite{bracher2011}) and the squared total momentum operator, $\langle \hat p^2 \rangle$, or their uncertainties, $\Delta \hat{r}$ and $\Delta \hat p$ (Ref.~\cite{khelashvili2022}) respectively, is expressed as an inequality in two and three dimensions,  and even in multidimensional spaces (Refs.~\cite{aljaber2016,Dehesa_2021,dehesa2021}), where $p$ represents the total momentum of the system and $r$ denotes the radial position. We will see for d-dimensional case, there will be an extra constant term in $\hat p_r$ does not affect $[\hat{r},\hat p_{r}]=i\hbar$, so we can make the radial uncertainty principle, $\Delta \hat{r}\Delta \hat p_r\geq \frac{\hbar}{2}$ for any arbitrary dimensional position space as in three dimension as well.

\section{Formulation for Radial uncertainty product}
The time-independent Schrodinger equation of a quantum particle of mass $\mu$ in d-dimensional spherical (hyper-spherical) coordinate ($r,\theta_1, \theta_2,...\theta_{d-2},\phi$) is given by
\begin{equation*}
    - \frac{\hbar^2}{2\mu} \nabla^2 \psi+V(r)\psi=E\psi
\end{equation*}
where $\psi=\psi(r,\theta_1, \theta_2,...\theta_{d-2},\phi)\equiv \psi(r,\Omega)$.
As we are interested with d-dimensional Hydrogen atom, then obviously $V(r)$ is the hyper-spherically symmetric potential and the Laplacian in hyper-spherical coordinate (Ref.~\cite{smirnov2019}) is given by
\begin{equation*}
    \nabla^2=\frac{1}{r^{d-1}}\frac{\partial}{\partial{r}}\left(r^{d-1}\frac{\partial}{\partial{r}}\right)+\frac{1}{r^2}\nabla^2_{S^{d-1}}
\end{equation*}
where $\nabla^2_{S^{d-1}}$ is the Laplace–Beltrami operator (hyper-spherical Laplacian) on the $(d - 1)$ sphere with unit radius. A simple derivation of it is attached in Appendix A. The symbol for hyper-spherical Laplacian, we have used $\nabla^2$ here, while $\Delta$ is used by a wide range of authors. The Laplace–Beltrami operator on the sphere is directly related to the angular momentum squared operator, $\hat L^2$ (Ref.~\cite{trinhammer2012}) by
\begin{equation*}
    \hat L^2=-\hbar^2 \nabla^2_{S^{d-1}}
\end{equation*}
The compact form for radial derivative in the Laplacian can be rewritten equivalent in terms of two radial derivatives as
\begin{equation*}
    \frac{1}{r^{d-1}}\frac{\partial}{\partial{r}}\left(r^{d-1}\frac{\partial}{\partial{r}}\right)=\frac{\partial^2}{\partial r^2}+\frac{d-1}{r} \frac{\partial}{\partial r}
\end{equation*}
We can use a shorthand operator for this radial operator for our convenience, is given by
\begin{equation}
    \nabla_r^2=\frac{1}{r^{d-1}}\frac{\partial}{\partial{r}}\left(r^{d-1}\frac{\partial}{\partial{r}}\right)
\end{equation}
Now the Laplacian takes form of
\begin{equation}
    \nabla^2=\nabla_r^2-\frac{\hat L^2}{r^2\hbar^2}
\end{equation}
and then the Hamiltonian can be expressed as
\begin{equation}
    \hat H=- \frac{\hbar^2}{2\mu} \nabla_r^2 + \frac{\hat L^2}{2\mu r^2} +V(r)
\end{equation}
Now we want to extract the radial Schrodinger equation for our purpose of radial wave function leads to radial uncertainty product. We know for spherically symmetric potential the solution can be written as 
\begin{equation*}
    \psi(r,\Omega)=R(r)\times Y(\Omega)
\end{equation*}
where $R(r)$ is the radial wave function and $Y(\Omega)$ is hyper-spherical harmonics. Using this fact, the Schrodinger equation can be rewritten as
\begin{equation*}
    -\frac{\hbar^2}{2\mu} \frac{1}{R(r)}\nabla_r^2 R(r) + \frac{1}{2\mu} \frac{1}{Y(\Omega)} \hat L^2 Y(\Omega) + V(r) =E
\end{equation*}
Applying the method of separation of variables, the revised form of the above equation will be
\begin{equation*}
    \frac{1}{R(r)}\nabla_r^2 R(r)-\frac{2\mu}{\hbar^2} r^2V(r) + \frac{2\mu E}{\hbar^2} r^2 = \frac{1}{\hbar^2} \frac{1}{Y(\Omega)} \hat L^2 Y(\Omega) = \alpha^2
\end{equation*}
where $\alpha^2$ is the separation constant and two separated equations are given by
\begin{equation}
    \left[-\frac{\hbar^2}{2\mu}\nabla_r^2+V(r)+\frac{\alpha^2\hbar^2}{2\mu r^2}\right] R(r) = E\, R(r)
\end{equation}
\begin{equation}
     \hat L^2 Y(\Omega)= \alpha^2 \hbar^2 Y(\Omega)
\end{equation}
Clearly Eq. (4) is the radial Schrodinger equation but with unknown constant $\alpha^2$ and it can be determined from Eq. (5). The hyper-spherical harmonics are eigenfunctions
of the Laplace-Beltrami operator on $S^{d-1}$ with unit radius which satisfy the 
eigenvalue equation (Ref.~\cite{cohl2012}) is given by
\begin{equation*}
    \nabla^2_{S^{d-1}} Y^K_{\ell}(\Omega) = - \ell (\ell+d-2) Y^K_{\ell}(\Omega)
\end{equation*}
where $K$ stands for the remaining $(d-2)$ quantum numbers, are often called as generalized azimuthal quantum numbers and they are among the $d$ quantum numbers due to the $d$-dimensional space. Those two are $n$ and $\ell$ where $n$ is the principal quantum number ($n\in \mathbb{Z}^+$), $\ell$ (often used as $\ell_1$ in many literature) is the azimuthal quantum number ($\ell\in \{ 0,1,...,n-1\}$), is defined on $(d-1)$-dimensional subspace. Now using the relation $\hat L^2=-\hbar^2 \nabla^2_{S^{d-1}}$, we have the eigenvalue and the eigenvalue equation of $\hat L^2$ is given by
\begin{equation}
    \hat L^2 Y^K_{\ell}(\Omega)=\ell (\ell+d-2) \hbar^2 Y^K_{\ell}(\Omega)
\end{equation}
This equation gives us $\alpha^2=\ell (\ell+d-2)$ and which can be proved easily using the homogeneity and harmonicity characteristics of the hyper-spherical harmonics. The hyper-spherical harmonics are the restrictions of the homogeneous harmonic polynomials $\mathbb{P}(\textbf{x})$ of degree $\ell$ to the unit sphere $S^{d-1}$. We know that a harmonic polynomial satisfies the Laplace equation $\nabla^2_{\mathbb{R}^d}\mathbb{P}(\textbf{x})=0$ where
\begin{equation*}
    \nabla^2_{\mathbb{R}^d}\equiv \nabla^2=\frac{\partial^2}{\partial r^2}+\frac{d-1}{r} \frac{\partial}{\partial r} -\frac{\hat L^2}{r^2\hbar^2}
\end{equation*}
A homogeneous polynomial $\mathbb{P}(\textbf{x})$ of degree $\ell$ by definition, it satisfies $\mathbb{P}(\lambda\textbf{x})=\lambda^{\ell} \mathbb{P}(\textbf{x})$ for $\lambda$ is a scalar. Such a polynomial in hyper-spherical coordinate system can be written as $\mathbb{P}(r,\Omega)=r^{\ell}Y(\Omega)$. This separation reflects the radial scaling symmetry of the homogeneous polynomial. The Laplace equation upon it 
\begin{equation*}
    \left(\frac{\partial^2}{\partial r^2}+\frac{d-1}{r} \frac{\partial}{\partial r} -\frac{\hat L^2}{r^2\hbar^2}\right)(r^{\ell}Y(\Omega))=0
\end{equation*}
gives the Eq. (6). The Eq. (4) can be written as
\begin{equation}
    \left[\frac{\hat p_r^2}{2\mu}+V_{\text{eff}}(r)\right]R(r)=E\,R(r)
\end{equation}
where $\hat p_r$ is the radial momentum operator and $V_{\text{eff}}(r)$ is the effective potential. Clearly, we can derive $V_{\text{eff}}(r)$ formula if we can establish an relation between $\hat p_r^2$ and $\nabla^2_r$. That's okay, then what will be the radial momentum operator?
In classical mechanics, the radial momentum is defined by,
\begin{equation*}
    {p}_{rc}=\frac{1}{r}\vec{r}.\vec{p}
\end{equation*}
where $\vec r$ and $\vec{p}$ are radial vector and momentum vector respectively.
But in quantum mechanics, this definition becomes ambiguous, since the components of $\hat r$ and $\hat p$ don't commute. Therefore, $\hat p_{r}=\frac{1}{r}\hat r.\hat p$ is not Hermitian results it can't be an observable. To make $\hat p_{r}$ a Hermitian operator, we need to define newly a symmetric operator, is given by
\begin{equation*}
    \hat p_{r}=\frac{1}{2}\frac{1}{r}(\hat{r}.\hat{p}+\hat{p}.\hat{r})
\end{equation*}
where $\hat r=r\hat e_{r}$ and $\hat{p}$ is momentum operator, is defined by $\hat{p}=-i\hbar\Vec{\nabla}$.
Then the radial momentum operator (Refs.~\cite{jana2025,Paz2001}) is defined as
\begin{equation}
    \hat p_{r} R(r)= \frac{1}{2}(-i\hbar)[\hat e_{r}.\Vec{\nabla}R(r)+\Vec{\nabla}.(\hat e_{r}R(r))] 
\end{equation}
for an arbitrary radial wave function $R(r)$. For any dimensional spherically coordinate system, the divergence of the system must contain $\hat{e_{r}}\frac{\partial}{\partial{r}}$, which gives the first term of Eq. (8) as 
\begin{equation*}
    \hat e_{r}.\Vec{\nabla}R(r)=\frac{\partial R(r)}{\partial r}
\end{equation*}
The evaluation of the second term in Eq. (8) is bread and butter in two- and three-dimensional spherical coordinate systems, as we have explicit formulas for their divergence. However, for a 
d-dimensional spherical coordinate system, it becomes quite tough to go through the concept of explicit formula of the divergence. To handle this, we apply a different approach. The Gauss divergence theorem for any vector field, $\vec A$ is given by
\begin{equation}
    \int_V \vec{\nabla}.\vec{A} \,dV=\oint_S \vec{A}.d\vec{S}
\end{equation}
where in d-dimensional spherical coordinate system, the volume and surface elements are given by
\begin{equation*}
    dV=r^{d-1}drd\Omega \text{ and } d\vec{S}=\hat e_r R^{d-1} d\Omega
\end{equation*}
Then
\begin{equation*}
    \int_V \vec{\nabla}.\vec{A} \,dV=\int_0^R\int_{\Omega} \vec{\nabla}.\vec{A} \, r^{d-1}drd\Omega
\end{equation*}
Taking a purely radial vector field, $\vec A = A_r \hat r$,
\begin{equation*}
\begin{alignedat}{2}
    &\oint_S \vec{A}.d\vec{S}=\int_{\Omega} A_{r=R} \hat e_r.\hat e_r R^{d-1} d\Omega\\
    &=\int_0^R \frac{\partial}{\partial r} \left(r^{d-1}A_r\right) dr \int_{\Omega} d\Omega=\int_0^R\int_{\Omega} \frac{\partial}{\partial r} \left(r^{d-1}A_r\right) drd\Omega\\
\end{alignedat}
\end{equation*}
From Eq. (9),
\begin{equation*}
    \vec{\nabla}.\vec{A}=\frac{1}{r^{d-1}}\frac{\partial}{\partial r} \left(r^{d-1}A_r\right)
\end{equation*}
Now 
\begin{equation*}
    \Vec{\nabla}.(\hat e_{r}R(r))=\frac{1}{r^{d-1}} \frac{\partial}{\partial r} \left(r^{d-1}R(r)\right)=\frac{d-1}{r}R(r)+\frac{\partial R(r)}{\partial r}
\end{equation*}
It's quite easy to evaluate $\Vec{\nabla}.(\hat e_{r}R(r))$, if we work with d-dimensional Euclidean space rather than spherical (Ref.~\cite{Paz2001}). Using the vector identity 
\begin{equation*}
    \Vec{\nabla}.(\hat e_{r}R(r))=(\hat e_r.\vec{\nabla}) R(r) + R(r) \vec{\nabla}.\hat e_r
\end{equation*}
and $r^2=\sum_{i=1}^d x_i^2$, $\hat e_r=\frac{1}{r} \sum_{i=1}^d x_i \hat e_i$, i.e., $\vec{\nabla}.\hat e_r=\frac{d-1}{r}$, we have exactly same result.
Finally from Eq. (8), we can write the radial momentum operator formula in d-dimensional spherical coordinate system, is given by
\begin{equation}
    \hat p_r = -i\hbar \left(\frac{\partial}{\partial r}+\frac{d-1}{2r}\right)
\end{equation}
Quantum mechanically squaring this operator, we have the squared radial momentum operator, is given by
\begin{equation}
    \hat p^2_r=-\hbar^2 \left(\frac{\partial^2}{\partial r^2}+\frac{d-1}{r}\frac{\partial}{\partial r}+\frac{(d-1)(d-3)}{4r^2}\right)
\end{equation}
This operator is very important in the context of radial uncertainty product. Now we have reached to our main goal, establishing an relation between $\hat p_r^2$ and $\nabla^2_r$, which is given by
\begin{equation}
     \hat p_r^2=-\hbar^2 \left(\nabla_r^2+\frac{(d-1)(d-3)}{4r^2}\right)
\end{equation}
using this relation, we can rewrite the Eq. (4) as
\begin{equation}
\begin{alignedat}{2}
    &\left[\frac{\hat p_r^2}{2\mu}+\frac{\hbar^2}{2\mu}\frac{(d-1)(d-3)}{4r^2}+\frac{\hbar^2}{2\mu}\frac{\ell(\ell+d-2)}{r^2}+V(r)\right]R(r)\\
    &=E\,R(r)\\
\end{alignedat}
\end{equation}
and comparing it with Eq. (7), we have the effective potential formula (Ref.~\cite{aljaber1998}) is given by
\begin{equation}
V_{\text{eff}}(r)=V(r)+\frac{\hbar^2}{2\mu}\frac{\ell(\ell+d-2)}{r^2} + \frac{\hbar^2}{2\mu}\frac{(d-1)(d-3)}{4r^2}
\end{equation}
The probability density function for d-dimensional position space (Refs.~\cite{aljaber1998}) is given by 
\begin{equation}
    P(r)=r^{d-1}|R(r)|^2
\end{equation} 
The normalization condition is given by
\begin{equation*}
    \int_{0}^{\infty} P(r) dr = 1 \implies\int_{0}^{\infty} r^{d-1} |R(r)|^2 dr = 1
\end{equation*}
The expectation value of an operator $\hat{A}$ in position space is given by
\begin{equation*}
    \langle\hat{A}\rangle=\int_{0}^{\infty} r^{d-1} R^*(r)\hat{A}R(r) dr
\end{equation*}
Now the expectation values of $\hat{r}$, $\hat{r}^2$ are given by
\begin{equation}
    \langle \hat{r} \rangle=\int_{0}^{\infty} r^{d-1} R^*(r)\hat{r}R(r) dr=\int_{0}^{\infty} r^d |R(r)|^2 dr 
\end{equation}
\begin{equation}
    \langle \hat{r}^2 \rangle=\int_{0}^{\infty} r^{d-1} R^*(r^2)\hat{r}^2R(r) dr=\int_{0}^{\infty} r^{d+1} |R(r)|^2 dr
\end{equation}
and the expectation values of $\hat p_{r}$ and $\hat p^2_{r} $ are given by
\begin{equation}
    \langle \hat p_{r} \rangle=\int_{0}^{\infty} r^2R^*(r)(-i\hbar)\left(\frac{\partial}{\partial{r}}+\frac{d-1}{2r}\right)R(r) dr
\end{equation}
\begin{equation}
\begin{alignedat}{2}
     &\langle   \hat p_r^2  \rangle=\int_{0}^{\infty} r^2R^*(r)(-\hbar^2)\left(\frac{\partial^2}{\partial{r^2}}+\frac{d-1}{r}\frac{\partial}{\partial{r}}+\frac{(d-1)(d-3)}{4r^2}\right)\\
     &R(r) dr\\
\end{alignedat}
\end{equation}
By using Eq. (19), to find $\langle   \hat p_r^2  \rangle$ is a little bit lengthy, while there is a fantastic way to find $\langle\hat{p}^2_{r}\rangle$ using the concept of the total energy of the system and the effective potential. From Eq. (3) and using Eq. (12), we can write the Hamiltonian as
\begin{equation}
    \hat H= \frac{\hat p_r^2}{2\mu} +\frac{\hbar^2}{2\mu}\frac{(d-1)(d-3)}{4r^2}+ \frac{\hat L^2}{2\mu r^2} +V(r)
\end{equation}
The effective potential operator can defined using $\hat L^2$ instead of $\ell(\ell+d-2)\hbar^2$ in Eq. (14), which is given by
\begin{equation}
    \hat V_{\text{eff}}(r)=V(r)+\frac{\hat L^2}{2\mu r^2} + \frac{\hbar^2}{2\mu}\frac{(d-1)(d-3)}{4r^2}
\end{equation}
Then the Hamiltonian can be rewritten through the effective potential operator.
\begin{equation}
    \hat H= \frac{\hat p_r^2}{2\mu}+\hat V_{\text{eff}}(r)
\end{equation}
\begin{equation}
    \langle   \hat p_r^2  \rangle=2\mu (\langle   \hat H  \rangle - \langle  \hat V_{\text{eff}}(r)   \rangle)=2\mu (E - \langle  \hat V_{\text{eff}}(r)   \rangle)
\end{equation}
where $E$ is the energy eigenvalue of the system and the expectation value of the effective potential is given by
\begin{equation}
\begin{alignedat}{2}
    &\langle  \hat V_{\text{eff}}(r)   \rangle = \frac{\hbar^2}{2\mu}\left(\frac{(d-1)(d-3)}{4}+\ell(\ell+d-2)\right)\langle\frac{1}{r^2}\rangle\\
    &+\langle V(r)\rangle\\
\end{alignedat}
\end{equation}
By the definition of uncertainty, the uncertainties of radial position and radial momentum are given by 
\begin{equation}
\begin{aligned}
    \Delta \hat{r}=\sqrt{\langle \hat{r}^2 \rangle-\langle \hat{r} \rangle^2}\\
    \Delta \hat p_r=\sqrt{\langle\hat{p}^2_{r}\rangle-\langle\hat{p}_r\rangle^2}
\end{aligned}
\end{equation}
In a compact form, the radial momentum operator can represented as
\begin{equation*}
    \hat p_r=-i\hbar\left(\frac{1}{r}\frac{\partial}{\partial{r}}[r]-\frac{d-3}{2}\right)
\end{equation*}
Lets check whether $\hat{r}$ and $\hat{p_{r}}$ commute or not.
The commutator of  $\hat{r}$ and $\hat{p_{r}}$ is given by $[\hat{r},\hat{p_{r}}]=(\hat{r}\hat{p_{r}}-\hat{p_{r}}\hat{r})$ and considering $\hat p_r=-i\hbar\frac{1}{r}\frac{\partial}{\partial{r}}[r]$ upon any arbitrary radial wave function $R(r)$,
\begin{equation*}
[\hat{r},\hat{p_{r}}]R(r)=(-i\hbar)\left[\frac{\partial}{\partial{r}}[rR(r)]-\frac{1}{r}\frac{\partial}{\partial{r}}[r^2R(r)]\right] =i\hbar R(r)
\end{equation*}
For the extra constant term in $\hat p_r$ does not affect $[\hat{r},\hat p_{r}]=i\hbar$, so we can make the radial uncertainty principle, $\Delta \hat{r}\Delta \hat p_r\geq \frac{\hbar}{2}$ for any arbitrary dimensional position space. The one dimensional reflection is nothing but $\Delta \hat x\Delta \hat p_{x} \geq \frac{\hbar}{2}$ corresponding to ${[\hat{x},\hat{p}_{x}]=i\hbar}$.

\section{d-Dimensional Non-relativistic Hydro-genic Atom}
\subsection{Radial wave function}
The time-independent Schrodinger equation for the hydro-genic system is given by
\begin{equation}
   -\frac{\hbar^2}{2\mu}\nabla^2\psi(r,\Omega)-\frac{Zke^2}{r}\psi(r,\Omega)=E\psi(r,\Omega)
\end{equation}
where $Z$, $\mu$ is the atomic number and reduced mass of one electron system, $k=\frac{1}{4\pi\epsilon_{0}}$ and $e$ is the electronic charge. For $V(r)=-\frac{Zke^2}{r}$, the effective potential can be written as
\begin{equation}
    V_{\text{eff}}(r)=-\frac{Zke^2}{r}+\frac{\hbar^2}{2\mu}\frac{\ell(\ell+d-2)}{r^2} + \frac{\hbar^2}{2\mu}\frac{(d-1)(d-3)}{4r^2}
\end{equation}
The radial Schrodinger equation can be written as
\begin{equation}
\begin{alignedat}{2}
    &\left[-\frac{\hbar^2}{2\mu} \left(\frac{1}{r^{d-1}}\frac{d}{dr}\left(r^{d-1}\frac{d}{dr}\right)\right) -\frac{Zke^2}{r} + \frac{\ell(\ell + d-2)\hbar^2}{2\mu r^2}\right]\\
    &R(r)= E R(r)\\
\end{alignedat}
\end{equation}
The solution (Ref.~\cite{aljaber1998}) of it is given by
\begin{equation}
\begin{alignedat}{2}
    &R_{n_r\ell}(r)=N_{n_r\ell}e^{-{\frac{Zr}{\left(n_r+\ell+\frac{d-1}{2}\right)a_{0}}}}\\
    &\left(\frac{2Z}{\left(n_r+\ell+\frac{d-1}{2}\right)a_{0}}r\right)^\ell L^{2\ell+d-2}_{n_r}\left({\frac{2Z}{\left(n_r+\ell+\frac{d-1}{2}\right)a_{0}}r}\right)\\
\end{alignedat}
\end{equation}
where $n_r$ is radial quantum number, which is specifically used for the radial solution of the Schrodinger equation and $a_{0}$ is the Bohr radius.
The principal quantum number, $\nu$ ($n'$ in Ref.~\cite{aljaber1998}) for d-dimensional hydro-genic system can be defined by
\begin{equation*}
    \nu=n_r+\ell+\frac{d-1}{2}
\end{equation*}
We have the principal quantum number, $n$ (Refs.~\cite{BransdenJoachain1989,aljaber1998,jana2025}) of usual three dimensional hydro-genic system is defined through
\begin{equation*}
    n=n_r+\ell+1
\end{equation*}
The fact is that the principal quantum number for d-dimensional quantum system, $\nu$ can take non-integer value for even dimensional cases. To avoid this, we will use the usual three dimensional quantum number $n$, which can take any positive integer value, they are related by
\begin{equation}
    \nu=n+\frac{d-3}{2}
\end{equation}
because of $d=3$, this relation reflects $\nu \equiv n$. Then the revised radial wave function (Refs.~\cite{aljaber1998}) from Eq. (29) can be written as,
\begin{equation}
\begin{alignedat}{2}
    &R_{n\ell}(r)=N_{n\ell}\, e^{-{\frac{Zr}{\left(n+\frac{d-3}{2}\right)a_{0}}}}\left(\frac{2Z}{\left(n+\frac{d-3}{2}\right)a_{0}}r\right)^\ell \\
    &L^{2\ell+d-2}_{n-\ell-1}\left({\frac{2Z}{\left(n+\frac{d-3}{2}\right)a_{0}}r}\right)\\
\end{alignedat}
\end{equation}
where $N_{n\ell}$ is the normalization constant, it can be calculated by normalizing condition. Here $L$ stands for associated Laguerre function (Generalized Laguerre function)(Ref.~\cite{AbramowitzStegun1972}).
In the context of this paper, the three properties of the associated Laguerre function, have an important role, are given by\\
1. Orthonormal property 
\begin{equation}
    \int_{0}^{\infty}\rho^a e^{-\rho}L^a_b(\rho)L^a_c(\rho)d\rho=\frac{\Gamma{(a+b+1)}}{\Gamma{(b+1)}}\delta_{bc}
\end{equation}
Many authors like Griffith's (Ref.~\cite{Griffiths2018}) uses the orthogonality property as 
\begin{equation*}
    \int_{0}^{\infty}\rho^a e^{-\rho}L^a_b(\rho)L^a_c(\rho)d\rho=\frac{\Gamma{(a+b+1)}}{[\Gamma{(b+1)}]^3}\delta_{bc}
\end{equation*}
while this will not affect in our calculation (Ref.~\cite{jana2025}).\\
2. Recursive property
\begin{equation}
\begin{alignedat}{2}
    &\rho L^a_b(\rho)=(a+2b+1)L^a_b(\rho)\\
    &-\frac{b+1}{a+b+1}L^a_{b+1}(\rho)-(a+b)^2L^a_{b-1}(\rho)\\
\end{alignedat}
\end{equation} 
3. Derivative property
\begin{equation}
    \frac{d}{d\rho}L^a_b(\rho)=L^{a+1}_b(\rho)=\frac{1}{\rho}[bL^a_b(\rho)-(b+a)L^a_{b-1}(\rho)]
\end{equation}
Let's introduce a dimensionless parameter $\rho$ as 
\begin{equation}
    \rho=\frac{2Z}{\left(n+\frac{d-3}{2}\right)a_{0}}r
\end{equation}
With this substitution, the radial wave function takes the form
\begin{equation}
    R_{n\ell}(\rho)=N_{n\ell}e^{-\frac{\rho}{2}}{\rho}^{\ell}L^{2\ell+d-2}_{n-\ell-1}(\rho)
\end{equation}
To ensure proper normalization, we redefine the probability density as
\begin{equation}
    P(r)=\left({\frac{\left(n+\frac{d-3}{2}\right)a_{0}}{2Z}}\right)^{d-1}{\rho}^{d-1}|R_{n\ell}(\rho)|^2  
\end{equation}
Since the radial coordinate is transformed as in Eq. (35), the differential element transforms accordingly
\begin{equation}
    dr=\frac{\left(n+\frac{d-3}{2}\right)a_{0}}{2Z}d\rho
\end{equation}
This transformation allows us to express the normalization condition in terms of the dimensionless variable.
\begin{equation}
\begin{alignedat}{2}
    &\int_0^{\infty}P(r)dr\\ 
    &=\left({\frac{\left(n+\frac{d-3}{2}\right)a_{0}}{2Z}}\right)^{d} \int_0^{\infty}{\rho}^{d-1}|R_{n\ell}(\rho)|^2d\rho\\
    &=\left({\frac{\left(n+\frac{d-3}{2}\right)a_{0}}{2Z}}\right)^{d}|N_{n\ell}|^2\\
    &\int_0^{\infty}{\rho}\,e^{-\rho}{\rho}^{2\ell+d-2}\left[L^{2\ell+d-2}_{n-\ell-1}(\rho)\right]^2d\rho\\
\end{alignedat}
\end{equation}
We will use shorthand $I$ for large integral expression in our calculations and for this integral,
\begin{equation}
    I_1=\int_0^{\infty}{\rho}\,e^{-\rho}{\rho}^{2\ell+d-2}\left[L^{2\ell+d-2}_{n-\ell-1}(\rho)\right]^2d\rho
\end{equation}
Following similar trend, naming $2\ell+d-2=a$ and $n-\ell-1=b$, the integral $I_1$ becomes,
\begin{equation*}
\begin{alignedat}{2}
    &I_1\\
    &=\int_0^{\infty}\rho.e^{-\rho}\rho^a[L^a_b(\rho)]^2d\rho\\
    &=\int_0^{\infty}\rho^ae^{-\rho}\left[(a+2b+1)L^a_b(\rho)\right.\\
    &\left.-\frac{b+1}{a+b+1}L^a_{b+1}(\rho)-(a+b)^2L^a_{b-1}(\rho)\right]L^a_b(\rho)d\rho\\  
    &=(a+2b+1)\int_0^{\infty}\rho^ae^{-\rho}[L^a_b(\rho)]^2d\rho\text{ (using Eq. (32) )}\\
    &=(a+2b+1)\frac{\Gamma{(a+b+1)}}{\Gamma{(b+1)}}\\
\end{alignedat}
\end{equation*}
Using our nomenclature, we can recover $a+2b+1=2n+d-3$, $a+b+1=n+\ell+d-2$ and $b+1=n-\ell$. This gives us finally,
\begin{equation*}
I_1=(2n+d-3)\frac{\Gamma{(n+\ell+1)}}{\Gamma(n-\ell)}=(2n+d-3)\frac{(n+\ell)!}{(n-\ell-1)!}  
\end{equation*}
Here we are able to use factorial instead of Gamma function because of $n$, $\ell$ both are positive integers. Using $I_1$ in Eq. (39), we can easily achieve
\begin{equation}
    I_0=\frac{\Gamma{(a+b+1)}}{\Gamma{(b+1)}}=\frac{1}{|N_{n\ell}|^2}\frac{1}{2\left(n+\frac{d-3}{2}\right)}\left({\frac{\left(n+\frac{d-3}{2}\right)a_{0}}{2Z}}\right)^{-d}
\end{equation}
which plays a very important role in further calculations. Using the normalization condition, from Eq. (39), we can easily evaluate our normalization constant, is given by
\begin{equation}
N_{n\ell}=\sqrt{\left(\frac{2Z}{\left(n+\frac{d-3}{2}\right)a_{0}}\right)^d\frac{(n-\ell-1)!}{(2n+d-3)(n+\ell+d-3)!}}
\end{equation}
If we use the orthogonality property from Griffiths, then it will be different only in cubic term, is given by
\begin{equation*}
    N_{n\ell}=\sqrt{\left(\frac{2Z}{\left(n+\frac{d-3}{2}\right)a_{0}}\right)^d\frac{[(n-\ell-1)!]^3}{(2n+d-3)(n+\ell+d-3)!}}
\end{equation*}
The normalized radial wave function in d-dimension is given by
\begin{equation}
\begin{alignedat}{2}
    &R_{n\ell}(r)=\sqrt{\left(\frac{2Z}{\left(n+\frac{d-3}{2}\right)a_{0}}\right)^d\frac{(n-\ell-1)!}{(2n+d-3)(n+\ell+d-3)!}}\\
    &e^{-{\frac{Zr}{\left(n+\frac{d-3}{2}\right)a_{0}}}}\left(\frac{2Z}{\left(n+\frac{d-3}{2}\right)a_{0}}r\right)^\ell 
L^{2\ell+d-2}_{n-\ell-1}\left({\frac{2Z}{\left(n+\frac{d-3}{2}\right)a_{0}}r}\right)\\
\end{alignedat}
\end{equation}
We have plotted some normalized radial wave function and corresponding probability distribution function for different dimensions for fixed states in Figs. (1-4).
\begin{figure}[h!]
    \centering
    \includegraphics[width=0.5\textwidth]{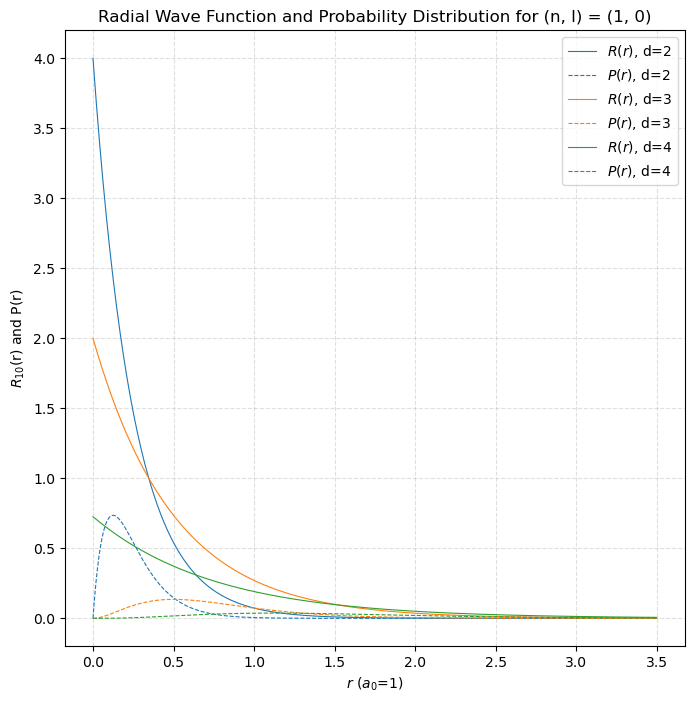}
    \caption{$R_{10}(r)$ and $P(r)$ vs $r$ with different $d$ for H-atom}
\end{figure}
\begin{figure}[h!]
    \centering
    \includegraphics[width=0.5\textwidth]{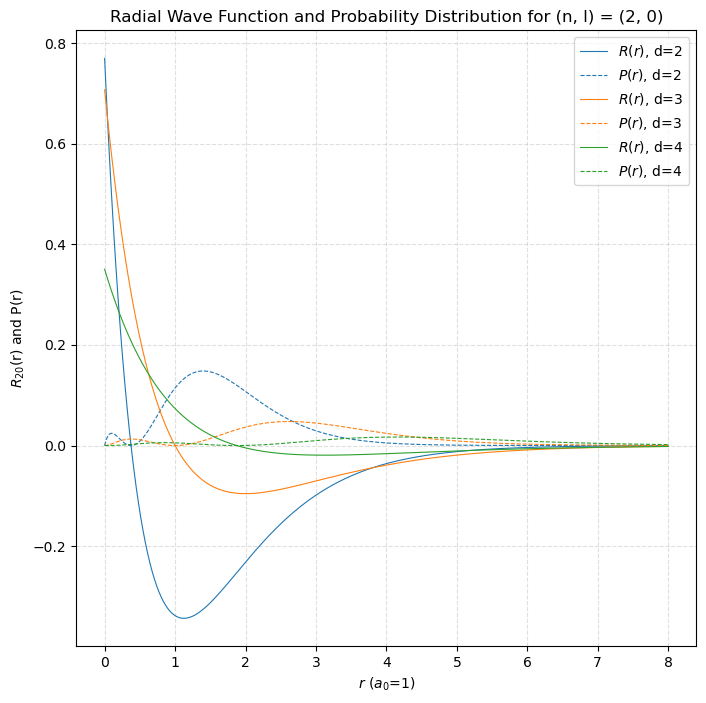}
    \caption{$R_{20}(r)$ and $P(r)$ vs $r$ with different $d$ for H-atom}
\end{figure}
\begin{figure}[h!]
    \centering
    \includegraphics[width=0.5\textwidth]{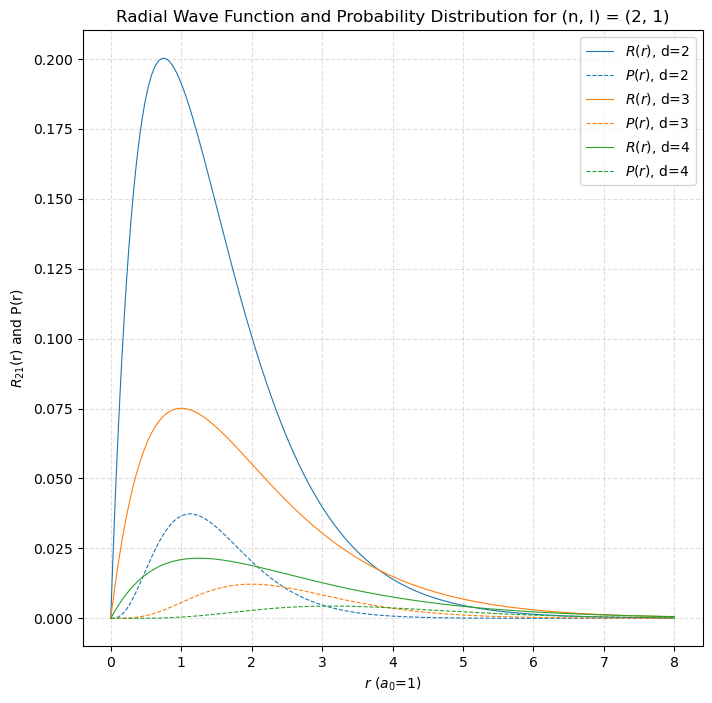}
    \caption{$R_{21}(r)$ and $P(r)$ vs $r$ with different $d$ for H-atom}
\end{figure}
\begin{figure}[h!]
    \centering
    \includegraphics[width=0.5\textwidth]{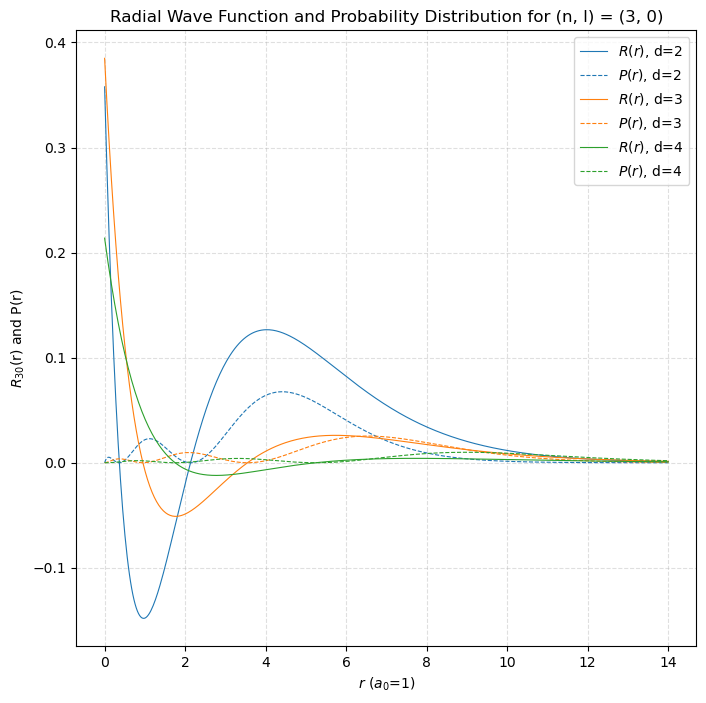}
    \caption{$R_{30}(r)$ and $P(r)$ vs $r$ with different $d$ for H-atom}
\end{figure}
\subsection{Uncertainty in Radial position}
According to Eq. (16),
\begin{equation}
\begin{alignedat}{2}
    &\langle \hat{r} \rangle\\
    &=\left({\frac{\left(n+\frac{d-3}{2}\right)a_{0}}{2Z}}\right)^{d+1} \int_0^{\infty}{\rho}^d |R_{n\ell}(\rho)|^2d\rho\\ 
    &=\left({\frac{\left(n+\frac{d-3}{2}\right)a_{0}}{2Z}}\right)^{d+1}|N_{n\ell}|^2 \\
    &\int_0^{\infty}{\rho}^2e^{-\rho}{\rho}^{2\ell+d-2}\left[L_{n-\ell-1}^{2\ell+d-2}(\rho)\right]^2d\rho\\
\end{alignedat}
\end{equation}
This integral is expressed in a shorthand as
\begin{equation*}
I_2=\int_0^{\infty}{\rho}^2e^{-\rho}{\rho}^{2\ell+d-2}\left[L_{n-\ell-1}^{2\ell+d-2}(\rho)\right]^2d\rho
\end{equation*}
Using our nomenclature, this integral becomes
\begin{equation*}
    I_2=\int_0^{\infty}{\rho}^2e^{-\rho}{\rho}^{a}\left[L_{b}^{a}(\rho)\right]^2d\rho
\end{equation*}
Using the recursive property of Eq. (33),
\begin{equation}
\begin{alignedat}{2}
    &\rho^2L^a_b(\rho)\\ 
    &=[(a+2b+1)^2+(b+1)(a+b+1)+b(a+b)]L_b^a(\rho)\\
    &+\left[-\frac{(a+2b+1)(b+1)}{a+b+1}-\frac{(b+1)(a+2b+3)}{a+b+1}\right]L_{b+1}^a(\rho)\\
    &+[-(a+2b+1)(a+b)^2-(a+2b-1)(a+b)^2]L_{b-1}^a(\rho)\\
    &+\frac{(b+1)(b+2)}{(a+b+1)(a+b+2)}L_{b+2}^a(\rho)\\
    &+(a+b)^2(a+b-1)^2L_{b-2}^a(\rho)\\
    &=\left[(a+2b+1)^2+(b+1)(a+b+1)+b(a+b)\right]L_b^a(\rho)\\
    &+\text{other terms contain } L_c^a(\rho) \text{ where } c\neq b \\
\end{alignedat}
\end{equation}
Using this relation and the orthogonal property of Eq. (32), the integral $I_2$ becomes
\begin{equation*}
\begin{alignedat}{2}
    &I_2=\int_{0}^{\infty}[(a+2b+1)^2+(b+1)(a+b+1)\\
    &+b(a+b)]z^a e^{-z}[L^a_b(\rho)]^2d\rho\\
    &=[(a+2b+1)^2+(b+1)(a+b+1)+b(a+b)]\\
    &\frac{\Gamma{(a+b+1)}}{\Gamma{(b+1)}}\\  
\end{alignedat}
\end{equation*}
Using our nomenclature, we can recover $(a+2b+1)^2=(2n+d-3)^2$, $(b+1)(a+b+1)=(n-\ell)(n+\ell+d-2)$ and $b(a+b)=(n-\ell-1)(n+\ell+d-3)$, leads $(a+2b+1)^2+(b+1)(a+b+1)+b(a+b)=d^2+d(6n-2\ell-7)+2(3n^2-9n-\ell^2+2\ell+6)$, which gives us finally,
\begin{equation*}
    I_2=[d^2+d(6n-2\ell-7)+2(3n^2-9n-\ell^2+2\ell+6)]\frac{\Gamma{(a+b+1)}}{\Gamma{(b+1)}}
\end{equation*}
Using this result with Eq. (41) and from Eq. (44), we have the formula of the expectation value of radial position in d-dimensional H-atom, is given by
\begin{equation}
    \langle \hat{r} \rangle= \frac{1}{4}\frac{a_{0}}{Z}[d^2+d(6n-2\ell-7)+2(3n^2-9n-\ell^2+2\ell+6)]
\end{equation}
The variation of $\langle r\rangle$ with dimensionality for different states are shown in Fig. (5).
\begin{figure}[h!]
    \centering
    \includegraphics[width=0.5\textwidth]{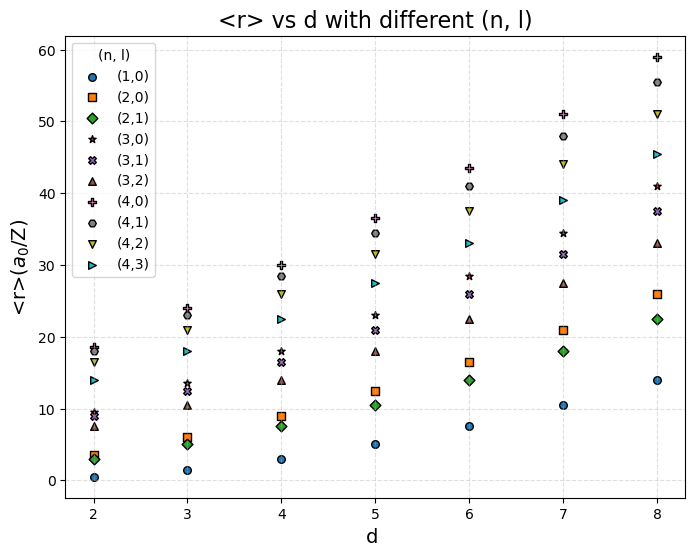}
    \caption{$\langle r\rangle$ vs $d$ with different $(n,\ell)$ for H-atom}
\end{figure}
Next, our objective is to evaluate $\langle \hat{r}^2 \rangle$, as from Eq. (17), 
\begin{equation}
\begin{alignedat}{2}
    &\langle \hat{r}^2 \rangle\\
    &=\left({\frac{\left(n+\frac{d-3}{2}\right)a_{0}}{2Z}}\right)^{d+2}\int_0^{\infty}{\rho}^{d+1}|R_{n\ell}(\rho)|^2d\rho\\ 
    &=\left({\frac{\left(n+\frac{d-3}{2}\right)a_{0}}{2Z}}\right)^{d+2} |N_{n\ell}|^2 \int_0^{\infty}{\rho}^3e^{-\rho}{\rho}^{2\ell+d-2}\left[L_{n-\ell-1}^{2\ell+d-2}(\rho)\right]^2d\rho\\
\end{alignedat}
\end{equation}
This integral is expressed in a shorthand as
\begin{equation*}
    I_3=\int_0^{\infty}{\rho}^3e^{-\rho}{\rho}^{2\ell+d-2}\left[L_{n-\ell-1}^{2\ell+d-2}(\rho)\right]^2d\rho
\end{equation*}
Using our nomenclature, this integral becomes
\begin{equation*}
    I_3=\int_0^{\infty}{\rho}^3 e^{-\rho}{\rho}^{a}\left[L_{b}^{a}(\rho)\right]^2d\rho
\end{equation*}
Rewriting Eq. (45) using some shorthands $U,V,W,X,Y$,
\begin{equation}
\begin{alignedat}{2}
    &\rho^2L^a_b(\rho)\\ &=U(a,b)L_b^a(\rho)+V(a,b)L_{b+1}^a(\rho)+W(a,b)L_{b-1}^a(\rho)\\
    &+X(a,b)L_{b+2}^a(\rho)+Y(a,b)L_{b-2}^a(\rho)\\
\end{alignedat}
\end{equation}
to focus on $\rho^3L^a_b(\rho)$, is given by
\begin{widetext}
\begin{equation*}
\begin{alignedat}{2}
    &\rho^3L^a_b(\rho)\\
    &=\rho\left[U(a,b)L_b^a(\rho)+V(a,b)L_{b+1}^a(\rho)+W(a,b)L_{b-1}^a(\rho)+X(a,b)L_{b+2}^a(\rho)+Y(a,b)L_{b-2}^a(\rho)\right]\\  
    &=U(a,b)\left[(a+2b+1)L_b^a(\rho)-\frac{b+1}{a+b+1}L_{b+1}^a(\rho)-(a+b)^2L_{b-1}^a(\rho)\right]\\
    &  +V(a,b)\left[(a+2b+3)L_{b+1}^a(\rho)-\frac{b+2}{a+b+2}L_{b+2}^a(\rho)-(a+b+1)^2L_b^a(\rho)\right]\\
    &  +W(a,b)\left[(a+2b-1)L_{b-1}^a(\rho)-\frac{b}{a+b}L_b^a(\rho)-(a+b-1)^2L_{b-2}^a(\rho)\right]\\
    &  +X(a,b)\left[(a+2b+5)L_{b+2}^a(\rho)+\frac{b+3}{a+b+3}L_{b+3}^a(\rho)-(a+b+2)^2L_{b+1}^a(\rho)\right]\\
    &  +Y(a,b)\left[(a+2b-3)L_{b-2}^a(\rho)-\frac{b-1}{a+b-1}L_{b-1}^a(\rho)-(a+b-2)^2L_{b-3}^a(\rho)\right]
\end{alignedat}
\end{equation*}
\end{widetext}
Now extracting the coefficients of $L_{b}^a(\rho)$, $\rho^3L^a_b(\rho)$ can be rewritten as
\begin{equation*}
\begin{alignedat}{2}
    &\rho^3L_b^a(\rho)\\
    &=\left[U(a,b)(a+2b+1)+V(a,b)[-(a+b+1)^2]\right.\\
    &\left.  +W(a,b)\left[-\frac{b}{a+b}\right]\right]L_b^a(\rho) \\
    &+\text{other terms not contain $L_c^a(\rho)$ where $c \neq b$}\\
    &=[U'(a,b)+V'(a,b)+W'(a,b)]L_b^a(\rho)\\
    &+ \text{other terms not contain $L_c^a(\rho)$ where $c \neq b$}\\
\end{alignedat}
\end{equation*}
where
\begin{equation*}
    U(a,b)=(a+2b+1)^2+(b+1)(a+b+1)+b(a+b)
\end{equation*}
\begin{equation*}
\begin{alignedat}{2}
    &U'(a,b)=U(a,b)(a+2b+1)=d^3+d^2(8n-2\ell-10)\\
    &+d(18n^2-2\ell^2-4n\ell-50n+10\ell+33)\\
    &+(12n^3-4n\ell^2-54n^2+6\ell^2+8n\ell+78n-12\ell-36)\\
\end{alignedat}
\end{equation*}
\begin{equation*}
    V(a,b)=-\frac{(a+2b+1)(b+1)}{a+b+1}-\frac{(b+1)(a+2b+3)}{a+b+1}
\end{equation*}
\begin{equation*}
\begin{alignedat}{2}
    &V'(a,b)=V(a,b)[-(a+b+1)^2]=2[d^2(n-\ell)\\
    &+d(3n^2-\ell^2-2n\ell-4n+4\ell)\\
    &+(2n^3-2n\ell^2-6n^2+2\ell^2+4n\ell+4n-4\ell)]\\
\end{alignedat}
\end{equation*}
\begin{equation*}
    W(a,b)=-(a+2b+1)(a+b)^2-(a+2b-1)(a+b)^2
\end{equation*}
\begin{equation*}
\begin{alignedat}{2}
    &W'(a,b)=W(a,b)\left[-\frac{b}{a+b}\right]=2[d^2(n-\ell-1)\\
    &+d(3n^2-\ell^2-2n\ell-10n+6\ell+7)\\
    &+(2n^3-2n\ell^2-12n^2+4\ell^2+4n\ell+22n-8\ell-12)]\\
\end{alignedat}
\end{equation*}
\begin{equation*}
\begin{alignedat}{2}
    &U'(a,b)+V'(a,b)+W'(a,b)=d^3+d^2(12n-6\ell-12)\\
    &+d(30n^2-6\ell^2-12n\ell-78n+30\ell+47)\\
    &+(20n^3-12n\ell^2-90n^2+18\ell^2+24n\ell+130n-36\ell-60)\\
\end{alignedat}
\end{equation*}
Using the orthogonal property of Eq. (32), the integral $I_3$ can be written as
\begin{equation*}
\begin{alignedat}{2}
    &I_3=\int_0^{\infty}[U'(a,b)+V'(a,b)+W'(a,b)]e^{-\rho}\rho^a[L^a_b(\rho)]^2d\rho\\  
    &=[U'(a,b)+V'(a,b)+W'(a,b)]\frac{\Gamma{(a+b+1)}}{\Gamma{(b+1)}}\\
\end{alignedat}
\end{equation*}
Using this result with Eq. (41) and from Eq. (47), we have the formula of the expectation value of squared radial position in d-dimensional H-atom, is given by
\begin{equation}
\begin{alignedat}{2}
   &\langle \hat{r}^2 \rangle=\frac{1}{8}.\frac{\left(n+\frac{d-3}{2}\right)a_0^2}{Z^2}[d^3+d^2(12n-6\ell-12)\\
   &+d(30n^2-6\ell^2-12n\ell-78n+30\ell+47)\\
   &+(20n^3-12n\ell^2-90n^2+18\ell^2+24n\ell+130n-36\ell-60)]\\
\end{alignedat}
\end{equation}
Borrowing the result of Eq. (46) for the expectation value of radial position in d-dimensional H-atom and using the result of Eq. (49), we can evaluate the generalized formula for uncertainty in radial position, which is given by
\begin{equation}
    \Delta \hat{r} =\frac{1}{4}\frac{a_0}{Z} \sqrt{
    \begin{aligned}
        &d^3(2n-2\ell-1) + d^2(6n^2-6\ell^2-18n+\\
        &12\ell+10) + d(8n^3-8\ell^3-36n^2+24\ell^2+62n\\
        &-22\ell-33) + (4n^4-4\ell^4-24n^3+16\ell^3\\
        &+62n^2-22\ell^2-78n+12\ell+36)\\
    \end{aligned}
    }
\end{equation}
The relative dispersion, also known as the coefficient of variation, in the measurement of the radial position is defined as 
\begin{equation*}
    \sigma_r = \frac{\Delta \hat{r}}{\langle \hat{r} \rangle}
\end{equation*}
which represents the ratio of the uncertainty to the expectation value of the radial position.  
The expression for the relative dispersion is given by
\begin{equation}
    \sigma_r=\frac{\sqrt{
    \begin{aligned}
        &d^3(2n-2\ell-1) + d^2(6n^2-6\ell^2-18n+\\
        &12\ell+10) + d(8n^3-8\ell^3-36n^2+24\ell^2+62n\\
        &-22\ell-33) + (4n^4-4\ell^4-24n^3+16\ell^3\\
        &+62n^2-22\ell^2-78n+12\ell+36)\\
    \end{aligned}
    }}{d^2+d(6n-2\ell-7)+2(3n^2-9n-\ell^2+2\ell+6)}
\end{equation}
The variation of the uncertainty  and the relative dispersion in radial position w.r.t. dimensionality are shown is Figs. (6,7).
\begin{figure}[h!]
    \centering
    \includegraphics[width=0.5\textwidth]{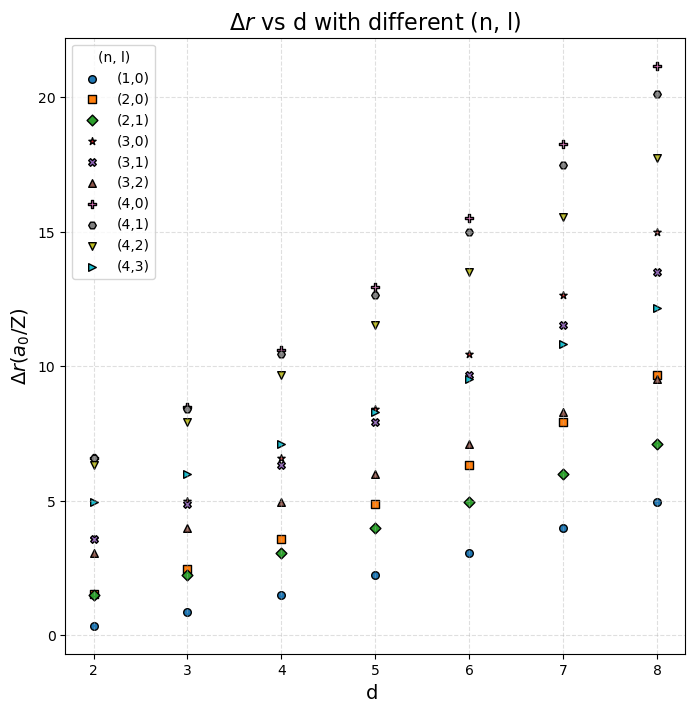}
    \caption{$\Delta r$ vs $d$ with different $(n,\ell)$ for H-atom}
\end{figure}
\begin{figure}[h!]
    \centering
    \includegraphics[width=0.5\textwidth]{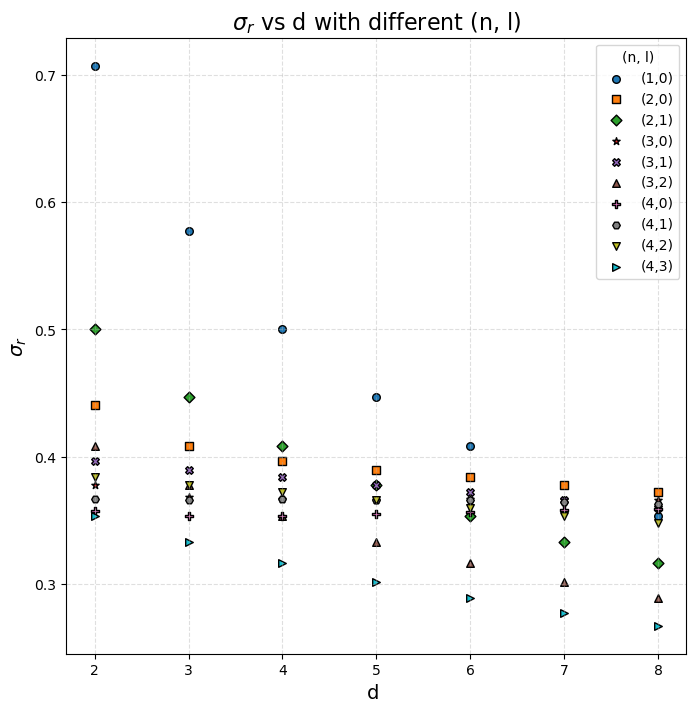}
    \caption{$\sigma_r$ vs $d$ with different $(n,\ell)$ for H-atom}
\end{figure}
\subsection{Uncertainty in Radial momentum}
We see that the radial momentum operator is given by,
\begin{equation*}
     \hat p_r=-i\hbar\left(\frac{1}{r}\frac{\partial}{\partial{r}}[r]-\frac{d-3}{2}\right)=-i\hbar\left(\frac{\partial}{\partial{r}}+\frac{d-1}{2r}\right)
\end{equation*}
According to Eq. (18),
\begin{equation}
\begin{alignedat}{2}
    &\langle \hat{p_{r}} \rangle\\
    &=\int_{0}^{\infty} r^{d-1} R_{n\ell}^*(r)(-i\hbar)\left(\frac{\partial}{\partial{r}}+\frac{d-1}{2r}\right)R_{n\ell}(r) dr\\
    &=-i\hbar\left[\int_0^{\infty}r^{d-1}R_{n\ell}\frac{\partial{R_{n\ell}}}{\partial{r}}dr+\frac{d-1}{2}\int_0^{\infty}r^{d-2}|R_{n\ell}|^2\,dr\right]\\
\end{alignedat}
\end{equation}
First integral of Eq. (52) becomes, 
\begin{equation*}
\begin{alignedat}{2}
   &I_4=\int_0^{\infty}r^{d-1}R_{n\ell}\frac{\partial{R_{n\ell}}}{\partial{r}}dr\\
   &=\left({\frac{\left(n+\frac{d-3}{2}\right)a_{0}}{2Z}}\right)^{d-1}\int_0^{\infty}{\rho}^{d-1}R_{n\ell}(\rho) \frac{\partial{R_{n\ell}}}{\partial{\rho}}d\rho\\
\end{alignedat}
\end{equation*}
Rewriting the radial wave function,
$R_{n\ell}(\rho)=N_{n\ell}e^{-\frac{\rho}{2}}{\rho}^{\ell}L^{a}_{b}(\rho)$, we have
\begin{equation*}
    \frac{dR_{n\ell}}{d\rho}=N_{n\ell}\left[-\frac{1}{2}e^{-\frac{\rho}{2}}{\rho}^{\ell}L_{b}^{a}(\rho)+e^{-\frac{\rho}{2}}{\ell}{\rho}^{\ell-1}L_{b}^{a}(\rho)+e^{-\frac{\rho}{2}}{\rho}^{\ell}\frac{dL_{b}^{a}(\rho)}{d\rho}\right]
\end{equation*}
Using the derivative property of Eq. (34),
\begin{equation}
\begin{alignedat}{2}
    &\frac{dR_{n\ell}}{d\rho}=N_{n\ell}\left[-\frac{1}{2}e^{-\frac{\rho}{2}}{\rho}^{\ell}L_b^a(\rho) \right. \\
    &\left. +(b+\ell)e^{-\frac{\rho}{2}}{\rho}^{\ell-1}L_b^a(\rho)-(b+a)e^{-\frac{\rho}{2}}{\rho}^{\ell-1}L_{b+1}^a(\rho)\right]\\
\end{alignedat}
\end{equation}
And as a result, $I_4$ becomes,
\begin{equation}
\begin{alignedat}{2}
    &I_4=\left({\frac{\left(n+\frac{d-3}{2}\right)a_{0}}{2Z}}\right)^{d-1}|N_{n\ell}|^2\\
    &\left[-\frac{1}{2}\int_0^{\infty}\rho.{\rho}^{a}e^{-\rho}[L_b^a(\rho)]^2d\rho\right.\\
    &+(b+\ell)\int_0^{\infty}{\rho}^{a}e^{-\rho}[L_b^a(\rho)]^2d\rho\\
    &\left.-(b+a)\int_0^{\infty}{\rho}^{a}e^{-\rho}L_b^a(\rho)L_{b-1}^a(\rho)d\rho\right]\\
\end{alignedat}
\end{equation}
The last integral will vanishes due to the orthogonality property and the first integral of $I_4$ is given by
\begin{equation}
\begin{alignedat}{2}
    &I_1=\int_0^{\infty}\rho.{\rho}^{a}e^{-\rho}[L_b^a(\rho)]^2d\rho\\  
    &=(a+2b+1)\int_0^{\infty}{\rho}^{a}e^{-\rho}[L_b^a(\rho)]^2d\rho\\
\end{alignedat}
\end{equation}
The second integral is one nothing but the orthonormal property of Eq. (32), is given by
\begin{equation}
    I_0=\int_0^{\infty}{\rho}^{a}e^{-\rho}[L_b^a(\rho)]^2d\rho
\end{equation}
Then expressing $I_1=(a+2b+1)I_0$, which leads
\begin{equation*}
\begin{alignedat}{2}
    &I_4=\left({\frac{\left(n+\frac{d-3}{2}\right)a_{0}}{2Z}}\right)^{d-1}|N_{n\ell}|^2\left[-\frac{a+2b+1}{2}+(b+\ell)\right]I_0\\
    &=-\left({\frac{\left(n+\frac{d-3}{2}\right)a_{0}}{2Z}}\right)^{d-1}|N_{n\ell}|^2\left(\frac{d-1}{2}\right)I_0
\end{alignedat}
\end{equation*}
Second integral of Eq. (52) becomes, 
\begin{equation*}
\begin{alignedat}{2}
    &I_5=\int_0^{\infty}r^{d-2}|R_{n\ell}|^2dr\\
    &=\left({\frac{\left(n+\frac{d-3}{2}\right)a_{0}}{2Z}}\right)^{d-1} |N_{n\ell}|^2\int_0^{\infty}e^{-\rho}{\rho}^{a}[L_{b}^{a}(\rho)]^2d\rho\\
    &=\left({\frac{\left(n+\frac{d-3}{2}\right)a_{0}}{2Z}}\right)^{d-1} |N_{n\ell}|^2 I_0\\
\end{alignedat}
\end{equation*}
The results of $I_4$ and $I_5$ clearly highlight
\begin{equation}
    \langle \hat p_r \rangle =0
\end{equation}
If we focus the integral $I_5$, it's nothing but $\langle \frac{1}{\hat r} \rangle$ and using Eq. (42), we have the expectation value of inverse radial position, is given by
\begin{equation}
    \langle \frac{1}{\hat r} \rangle = \frac{Z}{\left(n+\frac{d-3}{2}\right)^2a_0}
\end{equation}
which directly relates the expectation value of the potential of H-atom. For that purpose, let's focus on the energy eigenvalues. The energy eigenvalues (Refs.~\cite{aljaber1998}) are given by
\begin{equation}
    E_{\nu}=-\frac{\mu}{2\hbar^2} \frac{(Zke^2)^2}{\nu^2}
\end{equation}
and using Eq. (30) with the Bohr radius formula (Ref.~\cite{Zettili_2013}) $a_0=\frac{{\hbar}^2}{\mu ke^2}$, we can rewrite the energy eigenvalues as
\begin{equation}
    E_n=-\frac{1}{2\mu} \frac{Z^2\hbar^2}{\left(n+\frac{d-3}{2}\right)^2 a_0^2}
\end{equation}
The Virial theorem (Ref.~\cite{gupta2023}) states that for $V(r)\propto r^\alpha$, $\langle T  \rangle=\frac{\alpha}{2}\langle V  \rangle$. For Coulomb potential, $V(r)\propto \frac{1}{r}$ and $\langle T  \rangle=-\frac{1}{2}\langle V  \rangle$ that states that 
\begin{equation}
    \langle E  \rangle=\langle T  \rangle+\langle V  \rangle=\frac{1}{2}\langle V  \rangle
\end{equation}
For Hydrogen atom,
\begin{equation}
    V(r)=-\frac{Zke^2}{r}=-\frac{Z\hbar^2}{\mu a_0 r}
\end{equation}
By combining Eqs. (60–62), we can reproduce the result of Eq. (58). Next, our objective is to evaluate $\langle   \hat p_r^2  \rangle$ which includes a $\frac{1}{r^2}$ term. For that purpose, we aim to find $\langle\frac{1}{r^2}\rangle$ is given by
\begin{equation}
\begin{alignedat}{2}
    &\langle   \frac{1}{\hat{r}^2}  \rangle=\int_0^{\infty} r^{d-3}|R_{n\ell}|^2dr\\
    &=\left({\frac{\left(n+\frac{d-3}{2}\right)a_{0}}{2Z}}\right)^{d-2}|N_{n\ell}|^2\int_0^{\infty}\frac{1}{\rho}e^{-\rho}{\rho}^{a}\left[L_{b}^{a}(\rho)\right]^2d\rho\\
\end{alignedat}
\end{equation}
It's clear that this integral can't be done using the orthogonality condition and the recursive formula, we have applied before as a general prescription. Let redefine the integral as
\begin{equation}
    I_6(\lambda)=\int_0^{\infty}\frac{1}{\rho}e^{-\lambda\rho}{\rho}^{a}\left[L_{b}^{a}(\rho)\right]^2d\rho \text{ with } \lambda=1
\end{equation}
By differentiating this integral w.r.t. $\lambda$, we can redirect this integral to the integral $I_0$.
\begin{equation*}
\begin{alignedat}{2}
    &\frac{\partial I_6(\lambda)}{\partial \lambda}=-\int_0^{\infty}e^{-\lambda\rho}{\rho}^{a}\left[L_{b}^{a}(\rho)\right]^2d\rho\\
    &=-\lambda^{-(a+1)}\int_0^{\infty}e^{-\lambda\rho}{\lambda\rho}^{a}\left[L_{b}^{a}(\lambda\rho)\right]^2d(\lambda\rho)\\
    &=-\lambda^{-(a+1)}I_0=-\lambda^{-(a+1)}\frac{\Gamma{(a+b+1)}}{\Gamma{(b+1)}}\\
\end{alignedat}
\end{equation*}
Now integrating w.r.t. $\lambda$, we can recover our aimed integral, is given by
\begin{equation}
    I_6(\lambda)=\frac{\Gamma{(a+b+1)}}{a\Gamma{(b+1)}}\lambda^{-a} \implies I_6=\frac{\Gamma{(a+b+1)}}{a\Gamma{(b+1)}}
\end{equation}
Using Eq. (41), from Eq. (63), we have
\begin{equation}
    \langle \frac{1}{\hat r^2} \rangle = \frac{2Z^2}{a_0^2(2\ell+d-2)\left(n+\frac{d-3}{2}\right)^3} 
\end{equation}
An another approach (Ref.~\cite{Zettili_2013}) can be done taking the help of the radial Schrodinger equation. Consider
\begin{equation}
    f(r)=r^{\frac{d-1}{2}}R_{n\ell}(r)
\end{equation}
which gives us
\begin{equation}
    f''(r)=\frac{1}{4}r^{\frac{d-5}{2}}\left[(d^2-4d+3)R(r)+4r^2\left[\frac{(d-1)R'(r)}{r}+R''(r)\right]\right]
\end{equation}
and we have
\begin{equation}
    \nabla_r^2 R(r)=\frac{1}{r^{d-1}}\frac{\partial}{\partial{r}}\left(r^{d-1}\frac{\partial R(r)}{\partial{r}}\right)=\frac{(d-1)R'(r)}{r}+R''(r)
\end{equation}
By combining Eqs. (67–69), we can write
\begin{equation}
    \nabla_r^2 R(r)=\frac{1}{r^{\frac{d-1}{2}}}\left(f''(r)-\frac{d^2-4d+3}{4r^2}f(r)\right)
\end{equation}
The radial Schrodinger of Eq. (28) can be rewritten as
\begin{equation}
    -\frac{\hbar^2}{2\mu} \nabla_r^2 R(r) +\left[V(r) + \frac{\ell(\ell + d-2)\hbar^2}{2\mu r^2}\right]R(r)=E\,R(r)
\end{equation}
Using Eqs. (67,70) in Eq. (71) and rearranging bit,
\begin{equation}
    \frac{f''}{f}=-\frac{2\mu}{\hbar^2}(E-V(r))+\left[\frac{d^2-4d+3}{4}+\ell(\ell+d-2)\right]\frac{1}{r^2}
\end{equation}
Now differentiating this equation with respect to $\ell$,
\begin{equation}
    \frac{\partial}{\partial{\ell}}\left[\frac{f''}{f}\right]=\frac{2\ell+d-2}{r^2}-\frac{2\mu}{\hbar^2}\frac{\partial}{\partial \ell}(E-V(r))
\end{equation}
We keep $E-V(r)$ term here because of $n=n(\ell)=n_r+\ell+1$. Now multiplying both sides of Eq. (74) by $f^2(r)$ and integrating w.r.t. $r$ from $0$ to $\infty$,
\begin{equation}
\begin{alignedat}{2}
    &\int_0^{\infty}f^2\frac{\partial}{\partial{\ell}}\left[\frac{f''}{f}\right]dr=(2\ell+d-2)\int_0^{\infty}\frac{1}{r^2} f^2(r)dr\\
    &-\frac{2\mu}{\hbar^2}\int_0^{\infty}\frac{\partial}{\partial \ell}(E-V(r))f^2(r)dr\\
\end{alignedat}
\end{equation}
Using the fact $P(r)=f^2(r)$, let's rewrite Eq. (74),
\begin{equation}
    \int_0^{\infty}f^2\frac{\partial}{\partial{\ell}}\left[\frac{f''}{f}\right]dr=(2\ell+d-2)\langle   \frac{1}{\hat{r}^2}  \rangle-\frac{2\mu}{\hbar^2}\frac{\partial}{\partial \ell}\langle E-V(r)\rangle
\end{equation}
By using integration by parts, we can separate this integral to two integrals and again using integration by parts, we can show two integrals are equal, make the this integral zero. This is possible for $f(r=0)=0$ due to algebraic term and $f(r\to \infty)=0$ due to exponential term in radial wave function.
\begin{equation*}
    \int_0^{\infty}f^2\frac{\partial}{\partial{\ell}}\left[\frac{f''}{f}\right]dr=\int_0^{\infty} f \frac{\partial f''}{\partial \ell}dr-\int_0^{\infty} f''\frac{\partial f}{\partial \ell}dr=0
\end{equation*}
Finally, from Eq. (75), we have won an important formula, is given by
\begin{equation}
    \langle \frac{1}{\hat{r}^2} \rangle=\frac{2\mu}{\hbar^2} \frac{1}{2\ell+d-2} \frac{\partial}{\partial \ell}\langle E-V(r)\rangle
\end{equation}
This relation is an application of The Hellmann-Feynman theorem. We have discussed it in Appendix B. Now using Eq. (60,64) and the fact $\frac{\partial n}{\partial \ell}=1$, we can reproduce the result of Eq. (66). Now combining Eq. (23,24), we can write
\begin{equation}
\begin{alignedat}{2}
    &\langle \hat p_r^2 \rangle =2\mu (E-\langle V(r) \rangle)-\hbar^2\left[\frac{(d-1)(d-3)}{4}+\ell(\ell+d-2)\right]\\
    &\langle \frac{1}{\hat r^2} \rangle\\
\end{alignedat}
\end{equation}
For Hydrogen atom, using Eq. (61), we can rewrite
\begin{equation}
    \langle \hat p_r^2 \rangle =-\mu \langle V(r) \rangle-\hbar^2\left[\frac{(d-1)(d-3)}{4}+\ell(\ell+d-2)\right]\langle \frac{1}{\hat r^2} \rangle
\end{equation}
Again combining Eqs. (58,62),
\begin{equation}
    \langle V(r) \rangle = - \frac{Z^2\hbar^2}{\mu\left(n+\frac{d-3}{2}\right)^2a_0^2}
\end{equation}
Using Eqs. (66,79) in Eq. (78), we won the final result, is given by
\begin{equation}
    \langle \hat p_r^2 \rangle = \frac{Z^2\hbar^2}{\left(n+\frac{d-3}{2}\right)^2a_0^2} \left[1-\frac{(d-1)(d-3)+4\ell(\ell+d-2)}{(2n+d-3)(2\ell+d-2)}\right]
\end{equation}
Finally the uncertainty in radial momentum (not valid for $d=2$ with $\ell=0$) will be,
\begin{equation}
    \Delta \hat p_r=\frac{Z\hbar}{\left(n+\frac{d-3}{2}\right)a_0}\sqrt{1-\frac{(d-1)(d-3)+4\ell(\ell+d-2)}{(2n+d-3)(2\ell+d-2)}}
\end{equation}
The variation of the uncertainty of the radial momentum w.r.t. dimensionality is shown in Fig. (8).
\begin{figure}[h!]
    \centering
    \includegraphics[width=0.5\textwidth]{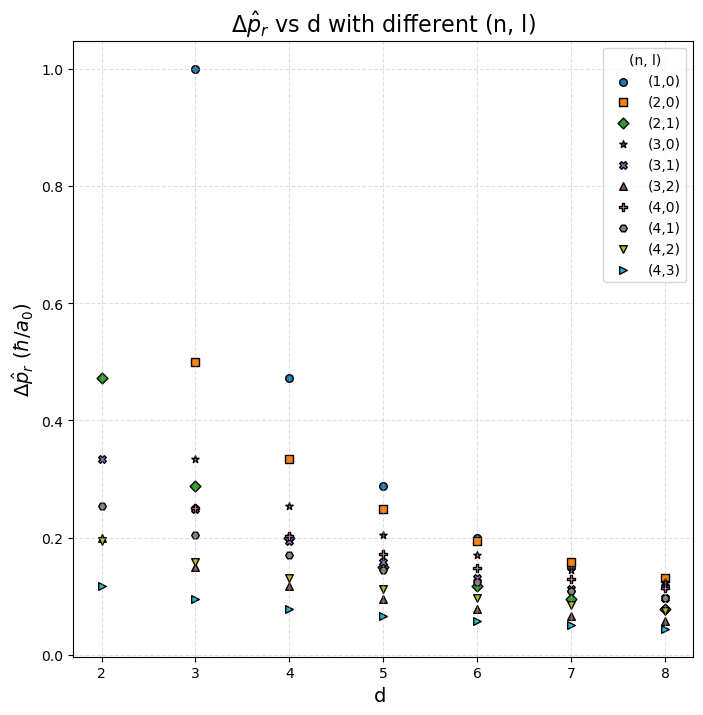}
    \caption{$\Delta \hat p_r$ vs $d$ with different $(n,\ell)$ for H-atom}
\end{figure}
\subsection{Radial Uncertainty Product}
Hence the radial uncertainty product (not valid for $d=2$ with $\ell=0$) will be,
\begin{equation}
\begin{alignedat}{2}
    &\Delta \hat{r} \Delta \hat p_r\\
    &= \sqrt{
    \begin{aligned}
        &d^3(2n-2\ell-1) + d^2(6n^2-6\ell^2-18n+\\
        &12\ell+10) + d(8n^3-8\ell^3-36n^2+24\ell^2+62n\\
        &-22\ell-33) + (4n^4-4\ell^4-24n^3+16\ell^3\\
        &+62n^2-22\ell^2-78n+12\ell+36)\\
    \end{aligned}
    }\\
    & \times\frac{\hbar}{4\left(n+\frac{d-3}{2}\right)}\sqrt{1-\frac{(d-1)(d-3)+4\ell(\ell+d-2)}{(2n+d-3)(2\ell+d-2)}}\\  
\end{alignedat}
\end{equation}
The radial uncertainty product w.r.t. dimensionality of Hydrogen atom problem is figured out in Fig. (9).
\begin{figure}[h!]
    \centering
    \includegraphics[width=0.5\textwidth]{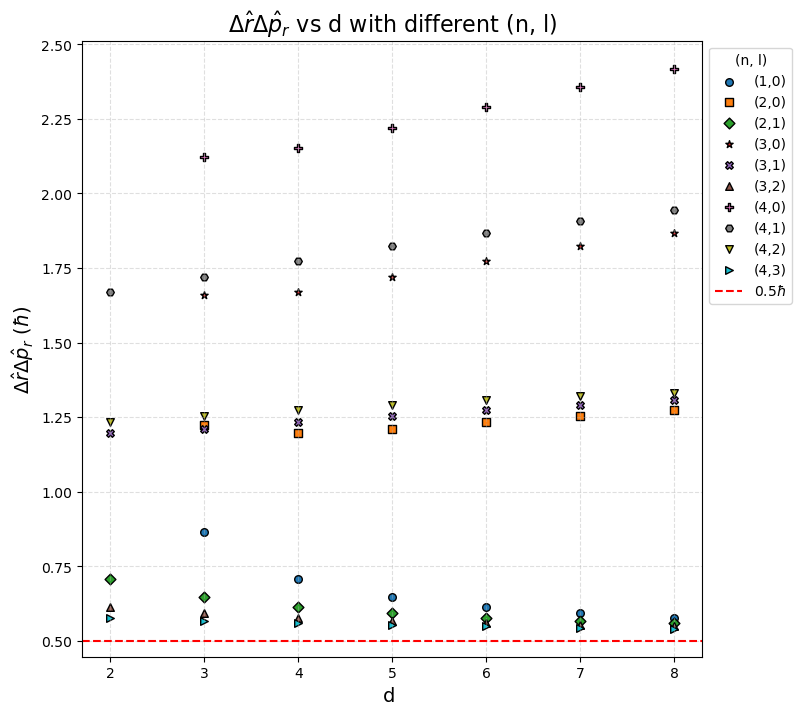}
    \caption{$\Delta \hat r \Delta \hat p_r$ vs $d$ with different $(n,\ell)$ for H-atom}
\end{figure}
\section{Lower Dimensional Results}
\subsection{General Results}
\begin{itemize}
    \item Radial probability distribution
    \begin{equation*}
    P(r)=r^{d-1}|R(r)|^2
\end{equation*}
    \begin{itemize}
        \item 2D: \begin{equation*}
    P(r)=r|R(r)|^2
\end{equation*}
        \item 3D: \begin{equation*}
    P(r)=r^{2}|R(r)|^2
\end{equation*}
    \end{itemize}
\end{itemize}
\begin{itemize}
    \item Expectation of an operator
    \begin{equation*}
    \langle\hat{A}\rangle=\int_{0}^{\infty} r^{d-1} R^*(r)\hat{A}R(r) dr
\end{equation*}
    \begin{itemize}
        \item 2D: \begin{equation*}
    \langle\hat{A}\rangle=\int_{0}^{\infty} r R^*(r)\hat{A}R(r) dr
\end{equation*}
        \item 3D: \begin{equation*}
    \langle\hat{A}\rangle=\int_{0}^{\infty} r^{2} R^*(r)\hat{A}R(r) dr
\end{equation*}
    \end{itemize}
\end{itemize}
\begin{itemize}
    \item Radial momentum operator
    \begin{equation*}
    \hat p_r = -i\hbar \left(\frac{\partial}{\partial r}+\frac{d-1}{2r}\right)
\end{equation*}
    \begin{itemize}
        \item 2D: \begin{equation*}
    \hat p_r = -i\hbar \left(\frac{\partial}{\partial r}+\frac{1}{2r}\right)
\end{equation*}
        \item 3D: \begin{equation*}
    \hat p_r = -i\hbar \left(\frac{\partial}{\partial r}+\frac{1}{r}\right)
\end{equation*}
    \end{itemize}
\end{itemize}
\begin{itemize}
    \item Eigenvalue equation of angular momentum squared operator
    \begin{equation*}
    \hat L^2 Y^K_{\ell}(\Omega)=\ell (\ell+d-2) \hbar^2 Y^K_{\ell}(\Omega)
\end{equation*}
    \begin{itemize}
        \item 2D: \begin{equation*}
    \hat L^2 Y^K_{\ell}(\Omega)=|\ell|^2  \hbar^2 Y^K_{\ell}(\Omega)
\end{equation*}
        \item 3D: \begin{equation*}
    \hat L^2 Y^K_{\ell}(\Omega)=\ell (\ell+1) \hbar^2 Y^K_{\ell}(\Omega)
\end{equation*}
    \end{itemize}
\end{itemize}
\begin{itemize}
    \item Effective potential operator
    \begin{equation*}
    \hat V_{\text{eff}}(r)=V(r)+\frac{\hat L^2}{2\mu r^2} + \frac{\hbar^2}{2\mu}\frac{(d-1)(d-3)}{4r^2}
\end{equation*}
    \begin{itemize}
        \item 2D: \begin{equation*}
    \hat V_{\text{eff}}(r)=V(r)+\frac{\hat L^2}{2\mu r^2} - \frac{\hbar^2}{2\mu}\frac{1}{4r^2}
\end{equation*}
        \item 3D: \begin{equation*}
    \hat V_{\text{eff}}(r)=V(r)+\frac{\hat L^2}{2\mu r^2}
\end{equation*}
    \end{itemize}
\end{itemize}
\begin{itemize}
    \item Radial momentum squared operator
    \begin{equation*}
    \hat p^2_r=-\hbar^2 \left(\frac{\partial^2}{\partial r^2}+\frac{d-1}{r}\frac{\partial}{\partial r}+\frac{(d-1)(d-3)}{4r^2}\right)
\end{equation*}
    \begin{itemize}
        \item 2D: \begin{equation*}
    \hat p^2_r=-\hbar^2 \left(\frac{\partial^2}{\partial r^2}+\frac{1}{r}\frac{\partial}{\partial r}-\frac{1}{4r^2}\right)
\end{equation*}
        \item 3D: \begin{equation*}
    \hat p^2_r=-\hbar^2 \left(\frac{\partial^2}{\partial r^2}+\frac{2}{r}\frac{\partial}{\partial r}\right)
\end{equation*}
    \end{itemize}
\end{itemize}
\begin{itemize}
    \item Expectation value of inverse squared radial position
    \begin{equation*}
    \langle \frac{1}{\hat{r}^2} \rangle=\frac{2\mu}{\hbar^2} \frac{1}{2\ell+d-2} \frac{\partial}{\partial \ell}\langle E-V(r)\rangle
\end{equation*}
    \begin{itemize}
        \item 2D: \begin{equation*}
    \langle \frac{1}{\hat{r}^2} \rangle=\frac{\mu}{\hbar^2 |\ell|} \frac{\partial}{\partial \ell}\langle E-V(r)\rangle
\end{equation*}
        \item 3D: \begin{equation*}
    \langle \frac{1}{\hat{r}^2} \rangle=\frac{2\mu}{\hbar^2} \frac{1}{2\ell+1} \frac{\partial}{\partial \ell}\langle E-V(r)\rangle
\end{equation*}
    \end{itemize}
\end{itemize}
\begin{itemize}
    \item Expectation value of radial momentum squared operator
    \begin{equation*}
    \begin{alignedat}{2}
    &\langle \hat p_r^2 \rangle =2\mu (E-\langle V(r) \rangle)-\hbar^2\left[\frac{(d-1)(d-3)}{4}+\ell(\ell+d-2)\right]\\
    &\langle \frac{1}{\hat r^2} \rangle\\
 \end{alignedat}
\end{equation*}
    \begin{itemize}
        \item 2D: \begin{equation*}
    \langle \hat p_r^2 \rangle =2\mu (E-\langle V(r) \rangle)-\hbar^2\left[-\frac{1}{4}+|\ell|^2\right]\langle \frac{1}{\hat r^2} \rangle
\end{equation*}
        \item 3D: \begin{equation*}
    \langle \hat p_r^2 \rangle =2\mu (E-\langle V(r) \rangle)-\hbar^2\ell(\ell+1)\langle \frac{1}{\hat r^2} \rangle
\end{equation*}
    \end{itemize}
\end{itemize}
\subsection{Hydrogen Atom}
\begin{itemize}
    \item Normalized radial wave function
    \begin{equation*}
    \begin{alignedat}{2}
    &R_{n\ell}(r)=\sqrt{\left(\frac{2Z}{\left(n+\frac{d-3}{2}\right)a_{0}}\right)^d\frac{(n-\ell-1)!}{(2n+d-3)(n+\ell+d-3)!}}\\
    &e^{-{\frac{Zr}{\left(n+\frac{d-3}{2}\right)a_{0}}}}\left(\frac{2Z}{\left(n+\frac{d-3}{2}\right)a_{0}}r\right)^\ell 
L^{2\ell+d-2}_{n-\ell-1}\left({\frac{2Z}{\left(n+\frac{d-3}{2}\right)a_{0}}r}\right)\\
\end{alignedat}
\end{equation*}
    \begin{itemize}
        \item 2D: \begin{equation*}
    \begin{alignedat}{2}
    &R_{n\ell}(r)=\sqrt{\left(\frac{2Z}{\left(n-\frac{1}{2}\right)a_{0}}\right)^2\frac{(n-|\ell|-1)!}{(2n-1)(n+|\ell|-1)!}}\\
    &e^{-{\frac{Zr}{\left(n-\frac{1}{2}\right)a_{0}}}}\left(\frac{2Z}{\left(n-\frac{1}{2}\right)a_{0}}r\right)^{|\ell|} 
L^{2|\ell|}_{n-|\ell|-1}\left({\frac{2Z}{\left(n-\frac{1}{2}\right)a_{0}}r}\right)\\
\end{alignedat}
\end{equation*}
        \item 3D: \begin{equation*}
    \begin{alignedat}{2}
    &R_{n\ell}(r)=\sqrt{\left(\frac{2Z}{na_{0}}\right)^3\frac{(n-\ell-1)!}{2n(n+\ell)!}}\\
    &e^{-{\frac{Zr}{na_{0}}}}\left(\frac{2Z}{na_{0}}r\right)^\ell 
L^{2\ell+1}_{n-\ell-1}\left({\frac{2Z}{na_{0}}r}\right)\\
\end{alignedat}
\end{equation*}
    \end{itemize}
\end{itemize}
\begin{itemize}
    \item Energy eigenvalues
    \begin{equation*}
    E_n=-\frac{1}{2\mu} \frac{Z^2\hbar^2}{\left(n+\frac{d-3}{2}\right)^2 a_0^2}
\end{equation*}
    \begin{itemize}
        \item 2D: \begin{equation*}
    E_n=-\frac{1}{2\mu} \frac{Z^2\hbar^2}{\left(n-\frac{1}{2}\right)^2 a_0^2}
\end{equation*}
        \item 3D: \begin{equation*}
    E_n=-\frac{1}{2\mu} \frac{Z^2\hbar^2}{n^2 a_0^2}
\end{equation*}
    \end{itemize}
\end{itemize}
\begin{itemize}
    \item Expectation value of radial position
    \begin{equation*}
    \langle \hat{r} \rangle= \frac{1}{4}\frac{a_{0}}{Z}[d^2+d(6n-2\ell-7)+2(3n^2-9n-\ell^2+2\ell+6)]
\end{equation*}
    \begin{itemize}
        \item 2D: \begin{equation*}
    \langle \hat r \rangle= \frac{1}{2}\frac{a_{0}}{Z}[3n^2-3n-|\ell|^2+1]
\end{equation*}
        \item 3D: \begin{equation*}
    \langle \hat{r} \rangle= \frac{1}{2}\frac{a_{0}}{Z}[3n^2-\ell(\ell+1)]
\end{equation*}
    \end{itemize}
\end{itemize}
\begin{itemize}
    \item Expectation value of radial position squared
    \begin{equation*}
    \begin{alignedat}{2}
   &\langle \hat{r}^2 \rangle=\frac{1}{8}.\frac{\left(n+\frac{d-3}{2}\right)a_0^2}{Z^2}[d^3+d^2(12n-6\ell-12)\\
   &+d(30n^2-6\ell^2-12n\ell-78n+30\ell+47)\\
   &+(20n^3-12n\ell^2-90n^2+18\ell^2+24n\ell+130n-36\ell-60)]\\
\end{alignedat}
\end{equation*}
    \begin{itemize}
        \item 2D: \begin{equation*}
    \langle \hat r^2 \rangle =\frac{1}{8}\frac{a_0^2}{Z^2}(2n-1)[n(10n^2-15n+11)-3|\ell|^2(2n-1)-3]
\end{equation*}
        \item 3D: \begin{equation*}
    \langle \hat{r}^2 \rangle=\frac{1}{2}.\frac{n^2a_0^2}{Z^2}[5n^2-3\ell(\ell+1)+1]
\end{equation*}
    \end{itemize}
\end{itemize}
\begin{itemize}
    \item Expectation value of inverse radial position
    \begin{equation*}
    \langle \frac{1}{\hat r} \rangle = \frac{Z}{\left(n+\frac{d-3}{2}\right)^2a_0} \implies \langle V(r) \rangle = - \frac{Z\hbar^2}{\mu a_0}\langle \frac{1}{\hat r} \rangle
\end{equation*}
    \begin{itemize}
        \item 2D: \begin{equation*}
    \langle \frac{1}{\hat r} \rangle = \frac{Z}{\left(n-\frac{1}{2}\right)^2a_0}
\end{equation*}
        \item 3D: \begin{equation*}
    \langle \frac{1}{\hat r} \rangle = \frac{Z}{n^2a_0}
\end{equation*}
    \end{itemize}
\end{itemize}
\begin{itemize}
    \item Expectation value of inverse squared radial position
    \begin{equation*}
    \langle \frac{1}{\hat r^2} \rangle = \frac{2Z^2}{a_0^2(2\ell+d-2)\left(n+\frac{d-3}{2}\right)^3}
\end{equation*}
    \begin{itemize}
        \item 2D: \begin{equation*}
    \langle \frac{1}{\hat r^2} \rangle = \frac{Z^2}{a_0^2|\ell|\left(n-\frac{1}{2}\right)^3} \text{ for } |\ell|\neq0
\end{equation*}
        \item 3D: \begin{equation*}
    \langle   \frac{1}{\hat{r}^2}  \rangle=\frac{2Z^2}{(2\ell+1)n^3a_0^2}
\end{equation*}
    \end{itemize}
\end{itemize}
\begin{itemize}
    \item Uncertainty in radial position
    \begin{equation*}
    \Delta \hat{r} =\frac{1}{4}\frac{a_0}{Z} \sqrt{
    \begin{aligned}
        &d^3(2n-2\ell-1) + d^2(6n^2-6\ell^2-18n+\\
        &12\ell+10) + d(8n^3-8\ell^3-36n^2+24\ell^2+62n\\
        &-22\ell-33) + (4n^4-4\ell^4-24n^3+16\ell^3\\
        &+62n^2-22\ell^2-78n+12\ell+36)\\
    \end{aligned}
    }
\end{equation*}
    \begin{itemize}
        \item 2D: \begin{equation*}
    \Delta \hat{r} = \frac{1}{2\sqrt{2}}\frac{a_0}{Z}\sqrt{n[2n^2(n-2)+7n-5]-|\ell|^2(2|\ell|^2-1)+1}
\end{equation*}
        \item 3D: \begin{equation*}
    \Delta \hat{r} =\frac{1}{2}\frac{a_0}{Z}\sqrt{n^2(n^2+2)-[\ell(\ell+1)]^2}
\end{equation*}
    \end{itemize}
\end{itemize}
\begin{itemize}
    \item Uncertainty in radial momentum ($\Delta \hat p_r=\sqrt{\langle \hat p_r^2 \rangle}$)
    \begin{equation*}
    \Delta \hat p_r=\frac{Z\hbar}{\left(n+\frac{d-3}{2}\right)a_0}\sqrt{1-\frac{(d-1)(d-3)+4\ell(\ell+d-2)}{(2n+d-3)(2\ell+d-2)}}
\end{equation*}
    \begin{itemize}
        \item 2D: \begin{equation*}
    \Delta \hat p_r=\frac{Z\hbar}{\left(n-\frac{1}{2}\right)a_0}\sqrt{1-\frac{2|\ell|-\frac{1}{2|\ell|}}{2n-1}} \text{ for } |\ell|\neq0
\end{equation*}
        \item 3D: \begin{equation*}
    \Delta \hat p_r=\frac{Z\hbar}{na_0}\sqrt{1-\frac{2\ell(\ell+1)}{n(2\ell+1)}}
\end{equation*}
    \end{itemize}
\end{itemize}

\begin{acknowledgments}
I sincerely express my gratitude to the Indian Institute of Technology (IIT), Delhi, for granting me valuable access to various journal platforms. I also extend my heartfelt appreciation to my classmates, Molla Suman Rahaman and Tamal Majumdar, for their constant encouragement and support throughout my work. Additionally, I acknowledge the essential role of computational tools such as Wolfram Mathematica, Jupyter Notebook, and Overleaf in facilitating this work.
\end{acknowledgments}

\appendix

\section{Hyper Spherical Coordinate System}
The Laplacian operator in $d$-dimensional spherical coordinates (Ref.~\cite{campos2020}) can be derived starting from its general expression in curvilinear coordinates. In an arbitrary coordinate system $(q^1, q^2, \dots, q^d)$, the Laplacian is given by  
\begin{equation*}
\nabla^2 f = \frac{1}{\sqrt{g}} \sum_{i=1}^{d} \frac{\partial}{\partial q^i} \left( \frac{\sqrt{g}}{h_i^2} \frac{\partial f}{\partial q^i} \right),
\end{equation*}
where $g$ is the determinant of the metric tensor and $h_i$ are the scale factors associated with the coordinate system. In $d$-dimensional Euclidean space $\mathbb{R}^d$, spherical coordinates are defined as  
\begin{equation*}
(q^1, q^2, \dots, q^d) = (r, \theta_1, \theta_2, ..., \theta_{d-2}, \phi).
\end{equation*}
where $r$ is the radial coordinate (ranging from $0$ to $\infty$),  $\theta_1, \theta_2, \dots, \theta_{d-2}$ are the polar angles (ranging from $0$ to $\pi$), and $\phi$ is the azimuthal angle (ranging from $0$ to $ 2\pi$). The corresponding metric takes the form  
\begin{equation*}
ds^2 = dr^2 + r^2 d\theta_1^2 + r^2 \sin^2\theta_1 d\theta_2^2 + \dots + r^2 \prod_{k=1}^{d-2} \sin^2\theta_k d\phi^2
\end{equation*}
From this, the scale factors are given by $h_r = 1$, $h_{\theta_1} = r$, $h_{\theta_2} = r \sin\theta_1$, $h_{\theta_3} = r \sin\theta_1 \sin\theta_2$, and so on up to $h_\phi = r \prod_{k=1}^{d-2} \sin\theta_k$. The volume element is  
\begin{equation*}
\sqrt{g} = r^{d-1} \prod_{k=1}^{d-2} (\sin\theta_k)^{d-k-1}.
\end{equation*}
Substituting $q^1=r$, $h_r=1$ and $\sqrt{g}=r^{d-1}$ into the general Laplacian expression, the radial term simplifies to  
\begin{equation*}
\frac{1}{\sqrt{g}} \frac{\partial}{\partial q^1} \left( \frac{\sqrt{g}}{h_r^2} \frac{\partial f}{\partial q^1} \right) = \frac{1}{r^{d-1}} \frac{\partial}{\partial r} \left( r^{d-1} \frac{\partial f}{\partial r} \right)
\end{equation*}
The angular part, involving the unit $(d-1)$-dimensional sphere, leads to  
\begin{equation*}
\nabla^2_{S^{d-1}} = \frac{1}{\sin^{d-2} \theta_1} \frac{\partial}{\partial \theta_1} \left( \sin^{d-2} \theta_1 \frac{\partial}{\partial \theta_1} \right) + \frac{1}{\sin^2 \theta_1} \nabla^2_{S^{d-2}}
\end{equation*}
In similar trend, the angular part, expanding up to the last term for $\phi$, we obtain  
\begin{equation*}
\begin{alignedat}{2}
&\nabla^2_{S^{d-1}} = \frac{1}{\sin^{d-2} \theta_1} \frac{\partial}{\partial \theta_1} \left( \sin^{d-2} \theta_1 \frac{\partial f}{\partial \theta_1} \right) \\
&+ \frac{1}{\sin^2 \theta_1} \frac{1}{\sin^{d-3} \theta_2} \frac{\partial}{\partial \theta_2} \left( \sin^{d-3} \theta_2 \frac{\partial f}{\partial \theta_2} \right)\\
&+ \dots + \frac{1}{\sin^2 \theta_1 \sin^2 \theta_2 \dots \sin^2 \theta_{d-3}} \frac{\partial^2 f}{\partial \phi^2}\\
\end{alignedat}
\end{equation*}
Thus, the final form of the $d$-dimensional Laplacian in spherical coordinates is  
\begin{equation*}
\nabla^2 f = \frac{1}{r^{d-1}} \frac{\partial}{\partial r} \left( r^{d-1} \frac{\partial f}{\partial r} \right) + \frac{1}{r^2} \nabla^2_{S^{d-1}} f
\end{equation*}
\section{Hellmann-Feynman Theorem}
The Hellmann-Feynman theorem (Ref.~\cite{esteve2010}) states that if \( H(\lambda) \) is a parameter-dependent Hamiltonian and \( |\psi(\lambda)\rangle \) is a normalized eigenstate of \( H(\lambda) \) with eigenvalue \( E(\lambda) \), i.e., $H(\lambda) |\psi(\lambda)\rangle = E(\lambda) |\psi(\lambda)\rangle$, then
\begin{equation*}
    \frac{dE}{d\lambda} = \left\langle \psi(\lambda) \left| \frac{dH}{d\lambda} \right| \psi(\lambda) \right\rangle.
\end{equation*}
A simple proof is attached here. Differentiating both sides of $H(\lambda) |\psi(\lambda)\rangle = E(\lambda) |\psi(\lambda)\rangle$ w.r.t. \( \lambda \), we get
\begin{equation*}
       \frac{dH}{d\lambda} |\psi(\lambda)\rangle + H(\lambda) \frac{d}{d\lambda} |\psi(\lambda)\rangle = \frac{dE}{d\lambda} |\psi(\lambda)\rangle + E(\lambda) \frac{d}{d\lambda} |\psi(\lambda)\rangle
\end{equation*}
Taking the inner product with \( \langle \psi(\lambda) | \), we obtain
\begin{equation*}
\begin{alignedat}{2}
       &\langle \psi(\lambda) | \frac{dH}{d\lambda} | \psi(\lambda) \rangle + \langle \psi(\lambda) | H(\lambda) \frac{d}{d\lambda} | \psi(\lambda) \rangle \\
       &= \frac{dE}{d\lambda} \langle \psi(\lambda) | \psi(\lambda) \rangle + E(\lambda) \langle \psi(\lambda) | \frac{d}{d\lambda} | \psi(\lambda) \rangle\\
\end{alignedat}
\end{equation*}
Since \( |\psi(\lambda)\rangle \) is normalized, using \( \langle \psi(\lambda) | \psi(\lambda) \rangle = 1 \), we have
\begin{equation*}
\begin{alignedat}{2}
       &\frac{dE}{d\lambda} = \langle \psi(\lambda) | \frac{dH}{d\lambda} | \psi(\lambda) \rangle \\
       &+ 
       \underbrace{\langle \psi(\lambda) | H(\lambda) \frac{d}{d\lambda} | \psi(\lambda) \rangle - 
       E(\lambda) \langle \psi(\lambda) | \frac{d}{d\lambda} | \psi(\lambda) \rangle}_{\text{Zero}}\\
\end{alignedat}
\end{equation*}
The last two terms cancel because \( H |\psi\rangle = E |\psi\rangle \). Thus, the Hellmann-Feynman theorem is proved. The expectation values of inverse and inverse squared radial position can be evaluated using this beautiful theorem, in a very simple way. Define the parameter $\lambda$ as $\lambda=Zke^2$, the Hamiltonian of H-atom (Eq. (26)) and energy eigenvalues (Eq. (59)) can be written as
\begin{equation*}
    H(\lambda)=-\frac{\hbar^2}{2\mu}\nabla^2-\frac{\lambda}{r} \text{ i.e. } \frac{dH(\lambda)}{d\lambda}=-\frac{1}{r}
\end{equation*}
\begin{equation*}
    E_{n}(\lambda)=-\frac{\mu}{2\hbar^2} \frac{\lambda^2}{\left(n+\frac{d-3}{2}\right)^2} \text{ i.e. } \frac{E_n(\lambda)}{d\lambda}=-\frac{Z}{\left(n+\frac{d-3}{2}\right)^2a_0}
\end{equation*}
where $a_0=\frac{\hbar^2}{\mu ke^2}$ is used and now using the Hellmann-Feynman theorem,
\begin{equation*}
    \langle \frac{1}{\hat r} \rangle =-\frac{dE_n(\lambda)}{d\lambda}= \frac{Z}{\left(n+\frac{d-3}{2}\right)^2a_0}
\end{equation*}
Define the parameter $\lambda$ once again as $\lambda=\ell$, the Hamiltonian (Eq. (20)) is given by
\begin{equation*}
    \hat H= \frac{\hat p_r^2}{2\mu} +\frac{\hbar^2}{2\mu}\frac{(d-1)(d-3)}{4r^2}+ \frac{ \ell(\ell+d-2)}{2\mu r^2} +V(r)
\end{equation*}
\begin{equation*}
    \frac{d\hat H}{d\ell}=\frac{2\ell+d-2}{2\mu}\frac{1}{r^2}
\end{equation*}
\begin{equation*}
    \frac{dE_n(\ell)}{d\ell}=\frac{dE_n(\ell)}{dn}\frac{dn}{d\ell}=\frac{Z^2\hbar^2}{\mu \left(n+\frac{d-3}{2}\right)^3a_0^2} \text{ using } \frac{dn}{d\ell}=1
\end{equation*}
Once again using the Hellmann-Feynman theorem,
\begin{equation*}
    \langle \frac{1}{\hat r^2} \rangle =\frac{2\mu}{2\ell+d-2}\frac{dE_n(\ell)}{d\ell}=\frac{2Z^2\hbar^2}{(2\ell+d-2) \left(n+\frac{d-3}{2}\right)^3a_0^2}
\end{equation*}
Even the Hellmann-Feynman theorem is so powerful, it can prove the Virial theorem in just two lines.  Define the parameter $\lambda$ as $\lambda=\mu$, the Hamiltonian and the energy eigenvalues (Eq. (59)) are given by
\begin{equation*}
    H(\mu)=-\frac{\hbar^2}{2\mu}\nabla^2+V(r) \text{ i.e. } \frac{dH(\mu)}{d\mu}=-\frac{1}{\mu}(H-V)
\end{equation*}
\begin{equation*}
    E_n(\mu)=-\frac{\mu}{2\hbar^2} \frac{(Zke^2)^2}{\left(n+\frac{d-3}{2}\right)^2} \text{ i.e. } \frac{dE_n(\mu)}{d\mu}=\frac{1}{\mu}E_n
\end{equation*}
The Hellmann-Feynman theorem directly says
\begin{equation*}
    -\langle (H-V) \rangle = E \text{ i.e. } -\langle T \rangle = E
\end{equation*}
and using $\langle H \rangle =E$, we achieve $E=\frac{1}{2}\langle V \rangle$ and $-2\langle T \rangle=\langle V \rangle$, they are results of the Virial theorem for Hydrogen atom, independent of dimensionality.



\nocite{*}
\begin{thebibliography}{99}



\bibitem{jana2025}
Avoy Jana,
\emph{Radial Uncertainty Product for Spherically Symmetric Potential in Position Space},
arXiv:2501.14831 [quant-ph] (2025),
\url{https://arxiv.org/abs/2501.14831}

\bibitem{Paz2001}
Gil Paz,
\emph{On the connection between the radial momentum operator and the Hamiltonian in n dimensions},
European Journal of Physics, 22(4), 337 (2001),
doi:10.1088/0143-0807/22/4/308. \url{https://arxiv.org/abs/quant-ph/0009046}



\bibitem{khelashvili2022}
Anzor Khelashvili and Teimuraz Nadareishvili,
\emph{Generalized Heisenberg uncertainty relation in spherical coordinates},
International Journal of Modern Physics B, 36(15), 2250072 (2022),
doi:10.1142/S0217979222500722. \url{https://doi.org/10.1142/S0217979222500722}

\bibitem{bracher2011}
Christian Bracher,
\emph{Uncertainty relations for angular momentum eigenstates in two and three spatial dimensions},
American Journal of Physics, 79(3), 313--319 (2011),
doi:10.1119/1.3534840. \url{https://doi.org/10.1119/1.3534840}

\bibitem{aljaber2016}
Sami M. Al-Jaber,
\emph{Uncertainty Relations for Some Central Potentials in N-Dimensional Space},
Applied Mathematics, 7(6), March 2016,
doi:10.4236/am.2016.76047. \url{https://doi.org/10.4236/am.2016.76047}

\bibitem{Dehesa_2021}
J. S. Dehesa,
\emph{Spherical-Symmetry and Spin Effects on the Uncertainty Measures of Multidimensional Quantum Systems with Central Potentials},
Entropy, 23(5), 607 (2021),
doi:10.3390/e23050607. \url{https://doi.org/10.3390/e23050607}


\bibitem{dehesa2021}
J. S. Dehesa and D. Puertas-Centeno,  
\emph{Multidimensional hydrogenic states: position and momentum expectation values},  
Journal of Physics B: Atomic, Molecular and Optical Physics, 54(6), 065006 (2021),  
doi:10.1088/1361-6455/abcdee. \url{http://dx.doi.org/10.1088/1361-6455/abcdee}.








\bibitem{smirnov2019}
A. Smirnov,
\emph{View on N-dimensional spherical harmonics from the quantum mechanical Pöschl-Teller potential well},
arXiv:1901.06711 [math-ph] (2019),
\url{https://arxiv.org/abs/1901.06711}


\bibitem{cohl2012}
Howard S. Cohl and Ernie G. Kalnins,
\emph{Fundamental solution of the Laplacian in the hyperboloid model of hyperbolic geometry},
arXiv:1201.4406 [math-ph] (2012),
\url{https://arxiv.org/abs/1201.4406}


\bibitem{trinhammer2012}
O. L. Trinhammer and G. Olafsson,
\emph{The Full Laplace-Beltrami operator on U(N) and SU(N)},
arXiv:math-ph/9901002 (2012),
\url{https://arxiv.org/abs/math-ph/9901002}.



\bibitem{aljaber1998}
S. M. Al-Jaber,  
\emph{Hydrogen Atom in N Dimensions},  
International Journal of Theoretical Physics, 37, 1289–1298 (1998),  
doi:10.1023/A:1026679921970. \url{https://doi.org/10.1023/A:1026679921970}


\bibitem{BransdenJoachain1989}
B. Bransden and C. Joachain,
\emph{Introduction to Quantum Mechanics},
Wiley, New York, 1989.


\bibitem{AbramowitzStegun1972}
M. Abramowitz and I. A. Stegun,
\emph{Handbook of Mathematical Functions with Formulas, Graphs, and Mathematical Tables},
9th printing, Dover, New York, 1972. Chapter 22: Orthogonal Polynomials, 771--802.


\bibitem{Griffiths2018}
David J. Griffiths and Darrell F. Schroeter,
\emph{Introduction to Quantum Mechanics},
3rd ed., Cambridge University Press, Cambridge, 2018.



\bibitem{Zettili_2013}
Nouredine Zettili,
\emph{Quantum Mechanics: Concepts and Applications},
2nd ed., Wiley, Hoboken, NJ, 2013.


\bibitem{gupta2023}
A. B. Gupta,  
\emph{Fundamentals of Classical Mechanics},  
3rd Edition, Paperback,  
New Central Book Agency, 2023.


\bibitem{Supriadi_2019}
B. Supriadi, A. Harijanto, M. Maulana, Z. R. Ridlo, W. D. Wisesa, and A. Nurdiniaya,
\emph{The function of the radial wave of a hydrogen atom in the principal quantum numbers (n) 4 and 5},
Journal of Physics: Conference Series, 1211, 012052 (2019),
doi:10.1088/1742-6596/1211/1/012052.


\bibitem{campos2020}
L. M. B. C. Campos and M. J. S. Silva,  
\emph{On hyperspherical associated Legendre functions: the extension of spherical harmonics to \(N\) dimensions}, 
arXiv:2005.09603 (2020),  
\url{https://arxiv.org/abs/2005.09603}.

\bibitem{esteve2010}
J. G. Esteve, F. Falceto, and C. García Canal,  
\emph{Generalization of the Hellmann–Feynman theorem},  
Physics Letters A, 374(6), 819–822 (2010),  
doi:10.1016/j.physleta.2009.12.005. \url{http://dx.doi.org/10.1016/j.physleta.2009.12.005}.

\bibitem{yang1991analytic}
X. L. Yang, S. H. Guo, F. T. Chan, K. W. Wong, and W. Y. Ching,  
\emph{Analytic solution of a two-dimensional hydrogen atom. I. Nonrelativistic theory},  
Phys. Rev. A, 43(3), 1186–1196 (1991),  
doi:10.1103/PhysRevA.43.1186. \url{https://link.aps.org/doi/10.1103/PhysRevA.43.1186}.






\end{thebibliography}
\end{document}